\documentclass[12pt, draftclsnofoot, onecolumn]{IEEEtran}

\usepackage{setspace}
\usepackage[margin=1in]{geometry}

\usepackage{cite}
\usepackage{amsmath,amssymb,amsfonts}
\usepackage{mathtools}
\usepackage{algorithmic}
\usepackage{graphicx}
\usepackage{textcomp}
\usepackage{multirow,booktabs}
\usepackage{stfloats,boldline}
\usepackage{url}
\usepackage{tablefootnote}
\usepackage{tabto}
\usepackage{float}
\usepackage{arydshln}
\usepackage[ruled,vlined]{algorithm2e}
\usepackage{bm}
\usepackage{bbm}
\usepackage{enumitem}
\usepackage[caption=false]{subfig}

\usepackage{amsthm}

\theoremstyle{definition}

\newtheorem{definition}{Definition}
\newtheorem{theorem}{Theorem}
\newtheorem{lemma}{Lemma}

\usepackage{xcolor}

\def\BibTeX{{\rm B\kern-.05em{\sc i\kern-.025em b}\kern-.08em
    T\kern-.1667em\lower.7ex\hbox{E}\kern-.125emX}}

\newif\ifstatus
\statustrue

\newcommand{\px}[1]{{\bf p}_{{\rm x}_{#1}}}%
\newcommand{\pbx}[1]{{\bf p}_{{\bf x}_{#1}}}%
\newcommand{\qx}[1]{{\bf q}_{{\rm x}_{#1}}}%

\newcommand{\pxa}[2]{{\bf p}_{{\rm x}^{\rm #2}_{#1}}}%
\newcommand{\pbxa}[2]{{\bf p}_{{\bf x}^{\rm #2}_{#1}}}%
\newcommand{\qbxa}[2]{{\bf q}_{{\bf x}^{\rm #2}_{#1}}}%
\newcommand{\qbtxa}[2]{{\bf \Tilde{q}}_{{\bf x}^{\rm #2}_{#1}}}%

\newcommand{\rvbx}[1]{{\bf x}_{#1}}
\newcommand{\svbx}[1]{\bm{x}_{#1}}
\newcommand{\rvbu}[1]{{\bf u}_{#1}}
\newcommand{\rvbxu}{{\bf x}_{\bf u}}
\newcommand{\svbu}[1]{\bm{u}_{#1}}
\newcommand{\svbxu}{\bm{x}_{\bm{u}}}
\newcommand{\rvx}[1]{{\rm x}_{#1}}

\newcommand{\rvu}[1]{{\rm u}_{#1}}

\newcommand{\pin}[1]{p_{{\bf x}_{\rm in}^{\rm #1}}}

\newcommand{\rndvin}[1]{{\bf x}_{\rm in}^{\rm #1}}
\newcommand{\rndvint}[1]{{\bf x}_{\rm int}^{\rm #1}}
\newcommand{\rndvsp}[1]{{\bf x}_{\rm sp}^{\rm #1}}
\newcommand{\smpvin}[1]{\bm{x}_{\rm in}^{\rm #1}}
\newcommand{\smpvint}[1]{\bm{x}_{\rm int}^{\rm #1}}
\newcommand{\smpvsp}[1]{\bm{x}_{\rm sp}^{\rm #1}}

\newcommand{\wt}[1]{{\rm wt_H}(#1)}

\newcommand{\ind}[1]{\mathbbm{1}(#1)}
\newcommand{\indmap}[1]{\varrho_{\mathcal{X}_{\scriptscriptstyle \rm #1}}}
\newcommand{\iindmap}[1]{\varrho^{-1}_{\mathcal{X}_{\scriptscriptstyle \rm #1}}}

\newcommand{\mr}[1]{\mathbf{m}_{\rm #1}}%
\newcommand{\pc}[1]{\rho^{\scriptscriptstyle \rm #1}}%
\newcommand{\pcc}[1]{\rho^{\scriptscriptstyle \rm #1}_{\max}}%
\newcommand{\bss}{\mathbf{b}_{\rm succ}}%
\newcommand{\bsp}{\mathbf{b}_{\rm prog}}%

\newcommand{\A}[1]{\mathcal{A}_{\rm #1}}%
\newcommand{\XA}[1]{\mathcal{X}^{\rm #1}}%
\newcommand{\XAin}[1]{\mathcal{X}^{\rm #1}_{\rm in}}%
\newcommand{\gA}[1]{\mathbf{G}_{{\mathcal{A}}_{\scriptscriptstyle \rm #1}}}%

\newcommand{\XAE}[2]{\mathcal{X}^{\rm #1}_{\mathcal{E_{\rm #2}}}}%
\newcommand{\xA}[1]{\mathbf{x}^{\rm #1}}%
\newcommand{\smpxA}[1]{\bm{x}^{\rm #1}}%
\newcommand{\xAn}[2]{{\bf x}_{#2}^{\rm #1}}%

\newcommand{\EP}[1]{\Theta_{\rm EP}({#1})}%
\newcommand{\mmc}[1]{\Theta_{\rm MC}(#1)}
\newcommand{\pmmc}[1]{\Theta_{{\rm MC}, i}(#1)}
\newcommand{\mmcA}[2]{\Theta^{\rm #1}_{\rm MC}(#2)}
\newcommand{\pmmcA}[2]{\Theta^{\rm #1}_{{\rm MC}, i}(#2)}

\newcommand{\DKL}[2]{{\rm D}_{\rm KL}\left(\,{#1}\, \|\, #2\,\right)}%
\newcommand{\dc}{d_{\mathcal{C}}}
\newcommand{\dv}{d_{\mathcal{V}}}

\newcommand{\scrl}[1]{\scriptscriptstyle #1}

\begin{document}
\title{Thermodynamic characterization of message-passing: Framework and initial study}

\author{Kalana~G.~Abeywardena,~\IEEEmembership{Graduate Student Member,~IEEE}\\
    Stark~C.~Draper,~\IEEEmembership{Senior~Member,~IEEE}% <-this % stops a space
\thanks{This work was supported by the Natural Sciences and Engineering Research Council (NSERC) of Canada through a Discovery Research Grant. This work was presented in part at the 2026 IEEE International Symposium on Information Theory (ISIT). \emph{(Corresponding author:~Kalana~G.~Abeywardena).}}
\thanks{The authors are with the Department of Electrical and Computer Engineering, University of Toronto, Toronto, ON M5S 3GA, Canada (e-mail: kalana.abeywardena@mail.utoronto.ca;
 stark.draper@utoronto.ca).}% <-this % stops a space
\thanks{\textcopyright\ 2026 IEEE. All rights reserved, including rights for text and data mining, and training of artificial intelligence and similar technologies. Personal use is permitted, but republication/redistribution requires IEEE permission.} %<-this % stops a space
 \thanks{The published version of this article is available in
\emph{IEEE Journal on Selected Areas in Information Theory}, vol. 7, pp. 396-410, 2026, doi: 10.1109/JSAIT.2026.3722113.}
}

\maketitle

\begin{abstract}
We study the thermodynamic efficiency of message-passing decoders using the \emph{mismatch cost} framework. This framework quantifies the excess entropy generated when a system operates away from the thermodynamically optimal initial distribution. It provides a computable lower bound on irrecoverable energy loss. The decoder is modeled as a main decoding program with local message updates implemented as subroutines. Memory is treated as an external copying subsystem whose thermodynamic cost is excluded. Studying the $[7,4]$ Hamming code transmitted over a binary erasure channel, we show how the channel-induced initialization statistics and the step-wise evolution of the algorithmic state distribution during decoding determine the total mismatch cost. We further show that different thermodynamically optimal state distributions, representing different underlying physical implementations of the decoder, can produce different thermodynamic costs for identical decoding dynamics. Conditioning the input distribution on correctable and non-correctable erasure patterns further reveals distinct thermodynamic signatures. These results demonstrate that thermodynamic cost is jointly governed by decoding dynamics, input statistics, and hardware characteristics. This points to opportunities for thermodynamically-informed decoder design to realize energy-efficient physical implementations.
\end{abstract}

\IEEEpeerreviewmaketitle

%%%%%%%%%%%%%%%%%%%%%%%%%%%%%%%%%%%%%%%%%%%%%%%%%%%%%%%%%%%%%%%%%%%%%%%%%%%%%%%%%%%%%%%%%%%%%%%%
\section{Introduction}
\label{sec:1.0}
Managing energy consumption is now a central consideration in the design of computing systems. At the micro level, power must be controlled to satisfy design constraints and prevent circuit failure, while at the global level, the energy demands of modern data centers have become a major technological and societal concern~\cite{wolpert2024}. As computing hardware continues to scale toward smaller physical regimes, the microscopic physical effects that underlie computation play an important role in determining energy efficiency. This has motivated growing interest in understanding how the physical dynamics of computing substrates influence energy consumption during computation~\cite{thermcompute2019}. Biological systems further illustrate that traditional semiconductor-based computing operates far from fundamental physical limits~\cite{kempes2017_cellsefficient, horowitz2020_nonequib}. In terms of fundamental limits, since computing is a physical process, its energy requirements have long been studied within the framework of (stochastic) thermodynamics~\cite{udo2012, parrondo2015, wolpert2019}. In this paper, we extend that framework to message-passing error-correction decoders and explore the resulting thermodynamic insights.

Foundational work by Szilard~\cite{szilard1964}, Landauer~\cite{landauer1961, landauer1996minimal}, and Bennett~\cite{bennett1982} established that information is fundamentally physical and that computation is a physical process. Landauer showed that logically irreversible operations generate a minimum energy $k_B T \Delta S$, where $k_B$ is Boltzmann’s constant, $T$ is the environment temperature, and $\Delta S$ is the decrease in Shannon entropy of the system. This yields the well-known $k_B T \ln 2$ bound for erasing a single bit~\cite{landauer1996minimal}. Bennett demonstrated that logically reversible computations can, in principle, avoid this energy cost if no information is lost~\cite{bennett1982}. Toffoli and Fredkin~\cite{fredkin1982} introduced reversible logic gates, and Feynman~\cite{feynman2018} emphasized that careful engineering of computational trajectories allows energy dissipation to be made arbitrarily small.

These early results are sometimes called ``semi-formal''~\cite{wolpert2019} because they only account for the reversible heat term $k_B T \Delta S$, commonly referred to as the Landauer cost~\cite{landauer1961}. This term captures the minimum energy dissipated as heat when information is irreversibly erased during computation. In principle, if all intermediate computational states are retained, the computation can be reversed to recover the exact input from the output, resulting in $\Delta S = 0$~\cite{bennett1982}. However, logical reversibility does not imply \emph{thermodynamic reversibility}. At the physical level, hardware implementations may exhibit asymmetric state-transition dynamics, such as unequal transistor switching rates or energetically unfavorable reverse transitions, which dissipate additional unrecoverable energy during computation. Achieving true \emph{thermodynamic reversibility} therefore requires the physical dynamics to remain arbitrarily close to thermal equilibrium, demanding extremely slow operation and precise microscopic control~\cite{feynman2018}. In practice, finite-time computation drives systems away from equilibrium, producing irreversible energy and entropy production that semi-formal analyses do not capture~\cite{van2023speed}. 

To address these limitations, \emph{stochastic thermodynamics} introduces tools to quantify the excess entropy produced by systems that operate far from equilibrium and that interact with their surrounding environment. Unfortunately, an exact characterization requires knowledge of microscopic system dynamics, such as charge-movement rates and transistor switching dynamics, which are generally unavailable~\cite{van2015trajectory, van2023speed}. Further, many lower bounds on entropy production~\cite{van2023speed, jarzynski2012equalities, horowitz2020_nonequib} still require quite detailed system information. The \emph{mismatch cost} framework~\cite{wolpert2019, wolpert20_ccts} reduces the need for such detailed information by providing a computable lower bound on entropy production using only observable state-space dynamics~\cite{wolpert24_DTMC}.

Mismatch cost measures the additional entropy generated when the system is initialized away from the thermodynamically optimal distribution over states, referred to as the prior. The prior encodes which system initialization states are thermodynamically preferred for a given computation through their assigned probabilities, without requiring a detailed circuit-level physical model. Previous work has shown that even digital devices designed to minimize heat dissipation generate a nonzero amount of entropy when the actual input distribution differs from the optimal energy-loss minimizing prior~\cite{wolpert24_cct2algo, wolpert24_comchannel, wolpert25_RASP}. In particular, the mismatch cost framework has been applied both to modular systems, where computation is composed of interacting reusable computational units, and to periodic systems, where the same sequence of computational steps is executed repeatedly for different inputs, cf.~\cite{wolpert20_interactsys, wolpert21_interactsys, wolper23_DFA}. The class of modular and periodic systems naturally includes message-passing decoders.

The energy consumption of message-passing decoding algorithms is of interest, as much of the energy consumed in a digital telecommunication system is consumed in the receiver and is spent on iterative inference and decoding. Message-passing decoders are core routines in modern standards, e.g., 5G and emerging 6G systems~\cite{he2023message}. In message-passing decoding, each iteration applies a fixed sequence of variable-node and check-node updates defined on a bipartite Tanner graph connecting variable and check nodes. Each update acts as a modular subroutine whose input distribution is determined by channel observations and incoming messages. Repeated execution of this structured computation on a physical system induces state evolution that changes both the system's and its
coupled environment's entropy. Hence, message passing decoding both motivates the stochastic thermodynamic framework and is of great practical importance. Together, these two considerations motivate an algorithmic-level thermodynamic analysis of message-passing decoders using the mismatch-cost framework to establish a lower bound on the entropy production associated with irreversible decoding dynamics. 

To our knowledge, the mismatch-cost framework has not yet been applied to message-passing decoders. Prior work on decoder energy consumption has primarily focused on hardware- and graph-level models. Thompson’s VLSI framework~\cite{thompson1979} relates energy to circuit area and computation time. Blake and Kschischang~\cite{blake2017, blake2018} apply this framework to Tanner-graph decoders, characterizing scaling with block length and node degree. Complementary studies~\cite{grover2011, grover2012} analyze system-level trade-offs using node activity and interconnect cost, highlighting the architectural limits of iterative decoding. While these approaches quantify energy through physical layout and switching activity, they do not capture the thermodynamic cost of the algorithmic message-update sequence itself.

\subsection{Our contributions}
\label{subsec:1.A}
In this paper, we present an algorithmic-level thermodynamic analysis of message-passing decoders using the mismatch-cost framework. Our focus is on developing a computable lower bound on entropy production defined at the level of observable algorithmic states, rather than on quantities derived from detailed physical or circuit-level dynamics. Using the $[7,4]$ Hamming code over the binary erasure channel (BEC) as an illustrative example, we show how stochastic-thermodynamic tools can be applied to iterative decoding algorithms. Our main contributions:
\begin{itemize}
    \item The introduction of a modeling framework that represents a message-passing decoder as an interactive computational system, with a main program performing decoding, and modular subroutines for variable- and check-node updates.
    \item The introduction of a computationally efficient method for handling extremely large state-transition structures by decomposing the transition matrix into smaller blocks. Even for the $[7,4]$ Hamming code, the main program has over $4.2$ million states. This makes direct computation intractable.
    \item The quantification of mismatch cost under uniform and non-uniform priors. By modeling different underlying physical implementations of the decoder, we show how the resulting prior distributions affect the thermodynamic cost of the same computation. 
    \item The analysis of mismatch cost across classes of erasure patterns on the BEC, revealing input-dependent thermodynamic signatures even in small codes.
\end{itemize}

\subsection{Outline}
\label{subsec:1.B}
The rest of the paper is organized as follows. Section~\ref{sec:preliminaries} reviews key concepts of message-passing decoders and entropy production. It introduces properties of mismatch cost and how to estimate it for periodic machines. Section~\ref{sec:framework} presents our framework for analyzing mismatch costs in message-passing decoding algorithms. Section~\ref{sec:results} reports numerical results for mismatch costs in message-passing decoders. While the $[7,4]$ Hamming code provides a controlled, interpretable setting, our long-term aim is to scale the framework to practical LDPC decoders and industry-relevant hardware. We discuss limitations of this work and conclude in Section~\ref{sec:futurework}.

\subsection{Notation}
\label{subsec:1.C}
For $n \in \mathbb{Z}_{+}$, define $[n] \coloneqq \{1,\dots,n\}$ and ${[n]}_0 \coloneqq \{0\} \cup [n]$. Vectors are denoted by bold lowercase letters, $\mathbf{x}$, matrices by bold uppercase letters, $\mathbf{X}$, and sets by calligraphic letters, $\mathcal{S}$, with cardinality $|\mathcal{S}|$. We write $\ind{\cdot}$ for the indicator function and $\delta(x,\theta) = \ind{x=\theta}$ for the Kronecker delta.
For a finite state space $\mathcal{X}$ with probability simplex $\mathcal{P_X}$, probability mass functions are denoted by $p_{\rvx{}} \in \mathcal{P_X}$. Fix a bijection $\indmap{}:\mathcal{X}\to[|\mathcal{X}|]$ that indexes states, with inverse $\indmap{}^{-1}$. The vectorized PMF $\px{}$ is defined as $(\px{})_{\indmap{}(x)} = p_{\rvx{}}(x)$ for $x \in \mathcal{X}$. This gives a one-to-one correspondence between the function $p_{\rvx{}}$ and the vector $\px{}$. We write $\langle \cdot \rangle_{p_{\rvx{}}}$ to denote expectation with respect to $p_{\rvx{}}$. Following the standard notation used in physics, the Shannon entropy of $p_{\rvx{}}$ is $S(p_{\rvx{}}) \coloneqq -\sum_{x \in \mathcal{X}} p_{\rvx{}}(x)\ln p_{\rvx{}}(x)$.

For a linear block code defined by parity-check matrix $\mathbf{H} \in \mathbb{F}_2^{m \times n}$, we can represent the code using a Tanner graph $G = (\mathcal{V}, \mathcal{C}, \mathcal{E})$ with variable nodes $\mathcal{V} = \{v_1,\dots,v_n\}$, check nodes $\mathcal{C} = \{c_1,\dots,c_m\}$, and edges $\mathcal{E} = \{(v,c) : {\rm H}_{cv} \neq 0, c \in [m], v \in [n]\}$. Variable and check node neighborhoods are $\mathcal{N}_v \subseteq \mathcal{C}$, $\mathcal{N}_c \subseteq \mathcal{V}$, with sets of neighborhoods $\mathcal{N_V} \coloneqq \{\mathcal{N}_v : v \in [n]\}$ and $\mathcal{N_C} \coloneqq \{\mathcal{N}_c : c \in [m]\}$. Node degrees are $d_v = |\mathcal{N}_v|$, $d_c = |\mathcal{N}_c|$, with maximum $d_{\max} = \max\{\dv{},\dc{}\}$, where $\dv = \max_v d_v$, $\dc = \max_c d_c$. Each variable node $v \in \mathcal{V}$ has a channel observation $\mu_v^0 \in \mathcal{O}$, where $\mathcal{O}$ is the observation alphabet. Messages along Tanner-graph edges take values in an algorithm-dependent alphabet $\mathcal{M}$, distinct from $\mathcal{O}$. While separate alphabets could distinguish check-to-variable and variable-to-check messages, we use a single message set.

%%%%%%%%%%%%%%%%%%%%%%%%%%%%%%%%%%%%%%%%%%%%%%%%%%%%%%%%%%%%%%%%%%%%%%%%%%%%%%%%%%%%%%%%%%%%%%%%
\section{Preliminaries}
\label{sec:preliminaries}
This section establishes the foundational concepts for our analysis. It provides a concise summary of key ideas from prior work rather than a comprehensive treatment.  The reader can find more detailed discussion in~\cite{kolchinsky2017, wolpert2019, wolpert24_DTMC, wolpert24_cct2algo}. We begin with a brief overview of message-passing decoding in Sec.~\ref{subsec:2.A}, motivating the study of such algorithms within a thermodynamic framework. Next, Sec.~\ref{subsec:2.B} introduces the thermodynamic cost of computation in a general physical system. In Sec.~\ref{subsec:2.C}, we present the mismatch cost as a fundamental lower bound on entropy production for generic physical computations. Finally, Sec.~\ref{subsec:2.D} extends this analysis to periodic algorithmic computations, the focus of this work.

\subsection{Message-passing decoding on Tanner Graphs}
\label{subsec:2.A}
Message-passing decoders for graph-based codes implement the sum-product algorithm or its variants~\cite{frey1997}. They iteratively update messages along the edges of the Tanner graph, from variable nodes (encoded symbols) to check nodes (parity constraints) and back. Despite numerous variations, computations at each node are simple and in ``low-density codes", each node processes only a small set of incoming messages from the corresponding (small) set of adjacent nodes. This combination of simple local updates and sparse graph structure makes message-passing decoding a natural candidate for analysis within the stochastic thermodynamic framework.

\subsection{Thermodynamic cost of computation in a general physical system}
\label{subsec:2.B}
Consider a finite-state system with state space $\mathcal{X}$. Over a fixed time interval $[0,\tau]$, the system's initial distribution $p_{\rvx{0}}$ evolves to a final distribution $p_{\rvx{\tau}}$ according to $p_{[0,\tau]}(x | x')$, where
\begin{equation}
    p_{\rvx{\tau}}(x) = \sum_{x'\in\mathcal{X}} p_{[0,\tau]}(x|x')\, p_{\rvx{0}}(x').
    \label{eq:2.1}
\end{equation}

Using the index map $\indmap{}$, define the state transition matrix ${\bf G}_{[0,\tau]} \in \mathbb{R}^{|\mathcal{X}| \times |\mathcal{X}|}$ on the interval $[0, \tau]$, with entries $G_{\indmap{}(x)\indmap{}(x')} = p_{[0,\tau]}(x \,|\, x')$. By construction, each column $j \in [\,|\mathcal{X}|\,]$ satisfies $\sum_{i=1}^{|\mathcal{X}|} G_{ij} = 1$. While the matrix depends on both the start and end times of the transitions, we suppress this dependence for notational compactness and simply denote the matrix as ${\bf G}$. Then~\eqref{eq:2.1} is written as $\px{\tau} = {\bf G}\px{0}$, where $\px{0}, \px{\tau} \in \mathbb{R}^{|\mathcal{X}|}$. In the remainder of this section, we will treat probability distributions as vectors in $\mathbb{R}^{|\mathcal{X}|}$ for notational simplicity. 

We consider the system to be coupled to $R \geq 1$ heat reservoirs indexed by $r \in [R]$. Each reservoir is a sufficiently large thermal environment that can absorb energy without any appreciable change in temperature over the time scale $[0,\tau]$~\cite{wolpert2019}. As a result, energy is dissipated into microscopic degrees of freedom of the reservoirs, such as electron scattering and lattice vibrations, with negligible effect on macroscopic properties. Each reservoir is characterized by $\beta_r = 1/k_BT_r$ where $k_B = 1.38 \times 10^{-23}\,\rm{J/K}$ is the Boltzmann constant and $T_r$ is its temperature. This multi-reservoir structure naturally arises in engineered systems, where different components interact with distinct thermal environments or cooling pathways (e.g., processors and memory dissipating into separate thermal domains), which cannot be captured by a single-reservoir model.

The interaction between the system and its heat reservoirs is determined by the underlying physical coupling mechanism. We model the evolution over a discretized interval $[0, \tau]$ as a stochastic trajectory $\rvbx{[0, \tau]} = (\rvx{0}, \rvx{1}, \dots, \rvx{\tau})$, where $\rvx{i} \in \mathcal{X}$ denotes the (random) system state at time step $i$. Each transition $\rvx{i} \rightarrow \rvx{i+1}$ may induce an energy transfer to an active reservoir $r \in [R]$ at that step. For an initial state $x \in \mathcal{X}$, let $\Bar{Q}_r(x)$ denote the expected cumulative energy transferred from the system to reservoir $r$ over stochastic trajectories initialized at $\rvx{0} = x$. For a general initial distribution $\pbx{0}$, the total expected heat flow to reservoir $r \in [R]$ is obtained by averaging over initial states as $\langle \Bar{Q}_r\rangle_{\px{0}} := \sum_x p_{\rvx{0}}(x)\, \Bar{Q}_r(x)$.

This construction admits a natural interpretation in digital circuits. The state $\rvx{i} \in \mathcal{X}$ is the joint logical configuration of a collection of logic gates, where each logical state is a coarse-grained description of an underlying ensemble of physical microstates. The transition matrix $\bf G$ describes probabilistic transitions between these logical states during computation. Physically, each logical transition is implemented through CMOS transistor switching and charge movement across semiconductor regions. This results in energy dissipation via capacitor charging and discharging, transient currents, and leakage. This energy is exchanged with coupled thermal reservoirs such as the silicon substrate or heat sink. In this view, $\Bar{Q}_r(x)$ captures the cumulative energy flow into reservoir $r$ by the physical switching activity underlying the logical state transitions starting from $x$, and $\langle \Bar{Q}_r\rangle_{\px{0}}$ is the expected energy flow under initial distribution $\px{0}$.

The expected heat flow $\Bar{Q}_r(x)$ from the system to reservoir $r \in [R]$ induces an entropy change $\beta_r \Bar{Q}_r(x)$ in that reservoir over the interval $[0, \tau]$~\cite{wolpert2019}. Summing over all reservoirs, we define the state-dependent total entropy flow from a system initialized in state $x \in \mathcal{X}$ into the $R$ reservoirs as \(f(x) = \sum_r \beta_r \Bar{Q}_r(x)\). For an initial distribution \(\px{0}\), the expectation \(\langle f \rangle_{\px{0}} = \sum_r \beta_r \langle \Bar{Q}_r \rangle_{\px{0}}\) gives the \emph{expected total entropy flow} into all reservoirs, denoted by $\Theta_{\rm EF}(\px{0})$. The associated \emph{expected entropy production} over the interval $[0, \tau]$ is defined as follows.
\begin{definition}[Expected entropy production~\cite{wolpert2019}] \label{def:2}
    Consider the (stochastic) evolution of an initial state distribution $\px{0}$ under dynamics $\bf G$ over a time interval $[0,\tau]$ with $\tau > 0$. Let $\Theta_{\rm EF}(\px{0}) := \sum_{r=1}^{R} \beta_r \langle \Bar{Q}_r \rangle_{\px{0}}$ denote the expected entropy flow from the system into $R$ reservoirs during the evolution. The expected entropy production is defined as
    \begin{equation}
        \EP{\px{0}} = \Theta_{\rm EF}(\px{0}) - \Big[\,S(\px{0}) - S({\bf G}\px{0})\,\Big],
        \label{eq:2.4}
    \end{equation}
    where $\Delta S(\px{0}) = S(\px{0}) - S({\bf G}\px{0})$ is the change in entropy between the initial and final state distributions under ${\bf G}$, with $S(\cdot)$ denoting Shannon entropy.
\end{definition}
From a statistical mechanical perspective, $\EP{\pbx{0}}$ quantifies the irreversibly dissipated portion of $\Theta_{\rm EF}(\px{0})$, arising from non-equilibrium effects in the physical implementation~\cite{kolchinsky2017, wolpert2019}. This energy is dispersed into microscopic degrees of freedom of the reservoirs, with physical manifestations in finite-speed hardware operation, including capacitive losses and leakage currents~\cite{cmosdesign2021}. The energy thereby becomes unavailable for recovery as useful work. In contrast, the Landauer cost $\Delta S(\px{0})$ captures the change in Shannon entropy induced by the transformation $\mathbf{G}$. This arises from stochastic dynamics and effective state-space compression in logically irreversible operations that merge multiple initial states into fewer distinguishable outcomes. Thus, the Landauer cost represents the reversible contribution to $\Theta_{\rm EF}(\px{0})$ associated with logical state transitions.

The second law of thermodynamics implies $\Theta_{\rm EP}(\px{0}) \ge 0$~\cite{jarzynski2012equalities}, ensuring $\Theta_{\rm EF}(\px{0}) \ge \Delta S(\px{0})$. Moreover, as shown in~\cite{wolpert24_cct2algo, wolpert25_RASP}, the Landauer cost satisfies $0 \le \Delta S(\px{0}) \le \ln |\mathcal{X}|$. The lower bound corresponds to logically irreversible operations (e.g., bit erasure) that reduce the system’s entropy. The upper bound corresponds to the entropy of the uniform distribution over $\mathcal{X}$. 
In general, $\Theta_{\rm EP}(\px{0})$ is strictly positive. However, it can be made arbitrarily small if the system evolves sufficiently slowly relative to the timescales over which its internal degrees of freedom relax to thermal equilibrium with coupled reservoirs~\cite{feynman2018}. Such \emph{quasi-static} (or \emph{near-equilibrium}) processes satisfy $\Delta S(\px{0}) \approx \Theta_{\rm EF}(\px{0})$. Here, the entropy change in the system due to logical operations is nearly balanced by entropy flow into the reservoirs, allowing the evolution, effectively, to be reversible.
In contrast, far-from-equilibrium processes generate large entropy production, with $\Theta_{\rm EF}(\px{0}) \gg \ln |\mathcal{X}|$, as the system evolves too rapidly to remain near equilibrium. The resulting excess entropy is irreversibly dispersed into the reservoirs and cannot be recovered as useful work. Consequently, $\Theta_{\rm EP}(\px{0})$ dominates the total thermodynamic cost.

As noted in Sec.~\ref{sec:1.0}, exact evaluation of $\Theta_{\rm EP}(\px{0})$ requires the (generally unavailable) full knowledge of the system’s physical, micro-scale, dynamics~\cite{wolpert24_cct2algo}. Therefore, in the following, we focus on developing a lower bound on $\Theta_{\rm EP}(\px{0})$ based on the \emph{mismatch cost}~\cite{wolpert2019, wolpert24_cct2algo}. 

\subsection{Mismatch cost as a lower bound on entropy production}
\label{subsec:2.C}
We now introduce the mismatch cost framework, which provides a general lower bound on the total expected entropy production $\EP{\px{0}}$ in Definition~\ref{def:2}. For fixed stochastic dynamics specified by $\mathbf{G}$ and the associated state-dependent entropy flows $\beta_r\Bar{Q}_r(x)$ for each reservoir $r \in [R]$, $\EP{\px{0}}$ depends on the choice of the initial state distribution $\px{0}$. This motivates the identification of a thermodynamically optimal initialization for the given implementation.
\begin{definition}[Prior distribution~\cite{wolpert2019}]\label{def:3}
    The \emph{prior distribution} is an initial distribution that achieves the minimum entropy production over the probability simplex $\mathcal{P_X}$:
    \begin{equation}
        \qx{0} = \underset{\mathbf{r}_{\rvx{0}} \in \mathcal{P_X}}{\arg\min} ~ \EP{\mathbf{r}_{\rvx{0}}}.
        \label{eq:2.6}
    \end{equation}
\end{definition}
The prior $\qx{0}$ is determined by the transition structure $\bf G$ and the state-dependent entropy flows $\beta_r \Bar{Q}_r(x)$, which characterize the thermodynamics of the implementation. The map $\mathbf{G}$ generates stochastic trajectories starting from state $x$, while $\Bar{Q}_r(x)$ denotes their average energy dissipation from device-level loss mechanisms. The initial distribution $\pbx{0}$ reweights these trajectories in the ensemble defining $\EP{\pbx{0}}$ without altering their dynamics. The prior $\qx{0}$ is therefore minimizes entropy production for the given implementation~\cite{kolchinsky2017, wolpert2019, wolpert24_cct2algo}. We consider two complementary viewpoints. In the first, a fixed physical implementation with entropy flows $\beta_r \Bar{Q}_r(x)$ induces a natural prior $\qx{0}$ via Definition~\ref{def:3}. In the second, a designed prior $\qx{0}$ characterizes the class of hardware implementations for which it serves as the entropy-minimizing distribution. In both cases, the energetic cost due to $\pbx{0} \neq \qx{0}$ is quantified by the mismatch cost defined below.

\begin{definition}[Mismatch cost~\cite{wolpert2019}] \label{def:4}
    Let $\px{0}$ be the actual initial distribution and $\qx{0}$ the prior distribution defined in~\eqref{eq:2.6}. The mismatch cost is defined as
    \begin{align}
        \mmc{\px{0}} &= \DKL{\px{0}}{\qx{0}} - \DKL{{\bf G}\px{0}}{{\bf G}\qx{0}} \label{eq:2.7.1}\\
        &= \EP{\px{0}} - \EP{\qx{0}}.
        \label{eq:2.7.2}
    \end{align}
\end{definition}
The mismatch cost quantifies the reduction in KL divergence between the actual and prior distributions before and after evolution under $\mathbf{G}$. Equivalently, it represents the thermodynamic penalty for initializing away from $\qx{0}$. We formalize this in the following theorem.
\begin{theorem}[Entropy production decomposition~\cite{kolchinsky2017}]
\label{thm:1}
    For any initial distribution $\px{0} \in \mathcal{P_{X}}$,
    \begin{equation}
        \EP{\px{0}} = \mmc{\px{0}} + \Theta_{\rm res},
        \label{eq:2.5}
    \end{equation}
    where $\Theta_{\mathrm{res}} := \Theta_{\rm EP}(\qx{0})$ is the \emph{residual entropy production}.
\end{theorem}
The proof is given in~\cite{kolchinsky2017}. Theorem~\ref{thm:1} decomposes entropy production into two components: a residual term $\Theta_{\rm res}$, representing the minimum achievable entropy production for the process and intrinsic to the physical implementation, and a mismatch cost $\mmc{\px{0}}$, which depends on the deviation of $\px{0}$ from the prior $\qx{0}$. 
\begin{lemma}[Properties of mismatch cost~\cite{wolpert24_cct2algo}]\label{lem:1}
    For an initial distribution $\px{0} \in \mathcal{P_X}$:
    \begin{enumerate}[label=(\roman*)]
        \item $\mmc{\px{0}} \ge 0$, with equality if and only if $\px{0} = \qx{0}$\label{lem:1.1}
        \item $\EP{\px{0}} \ge \mmc{\px{0}}$, with equality when $\Theta_{\rm res}=\EP{\qx{0}} = 0$.\label{lem:1.2}
    \end{enumerate}
\end{lemma}
Lemma~\ref{lem:1} establishes the non-negativity of the mismatch cost and a universal lower bound on $\EP{\px{0}}$. Part~\ref{lem:1.1} of Lemma~\ref{lem:1} follows from the data-processing inequality, since $\mathbf{G}$ is a stochastic map and $\DKL{\px{0}}{\qx{0}} \ge \DKL{{\bf G}\px{0}}{{\bf G}\qx{0}}$. 
Part~\ref{lem:1.2} follows from $\Theta_{\mathrm{res}}\ge 0$ and the decomposition in Theorem~\ref{thm:1}. When $\Theta_{\rm res} = 0$, the process is thermodynamically reversible under $\qx{0}$. Thus, the mismatch cost fully accounts for $\EP{\pbx{0}}$, making it a tight bound.

Mismatch cost is particularly useful because, once $\qx{0}$ and the transition map $\mathbf{G}$ are specified, it can be evaluated without reference to microscopic implementation details, depending only on the induced stochastic dynamics and entropy flows~\cite{wolpert21_interactsys, wolpert22_DFA, wolpert24_cct2algo}. Mismatch cost isolates the component of dissipation arising purely from the mismatch between the initialization $\px{0}$ and the thermodynamically optimal prior $\qx{0}$. Our work focuses on this lower bound on the total expected entropy production and the ensuing implications for efficient computation.

\subsection{Mismatch cost of algorithmic computation}
\label{subsec:2.D}
We now specialize the mismatch-cost framework introduced in Sec.~\ref{subsec:2.C} to algorithmic computation on a physical processor~\cite{wolpert24_cct2algo, wolpert24_DTMC}. An algorithm $\mathcal{A}$ is defined as an ordered sequence of instructions, where some instructions may repeat due to loops and others may execute conditionally~\cite{wolpert24_cct2algo, wolpert24_DTMC}. 
Although the physical system executing $\mathcal{A}$ has many microscopic degrees of freedom (e.g., electronic states and charge configurations), we coarse-grain over algorithm-relevant variables such as the program counter $(\rho)$, input, output, and intermediate registers, sampled at discrete instruction boundaries~\cite{wolpert24_cct2algo}. This yields a coarse-grained description in terms of a random algorithmic state $\mathbf{x}_i \in \mathcal{X}^{\mathcal{A}}$ at step $i$, where $\mathcal{X}^{\mathcal{A}}$ denotes the corresponding state space. The dynamics are then specified by a transition map $G_{\mathcal{A}} : \mathcal{X}^{\mathcal{A}} \to \mathcal{X}^{\mathcal{A}}$, such that $\mathbf{x}_{i+1} = G_{\mathcal{A}}(\mathbf{x}_i)$.

Following Sec.~\ref{subsec:2.C}, the dynamics are represented by a stochastic transition matrix $\gA{} \in \mathbb{R}^{|\mathcal{X}^{\mathcal{A}}| \times |\mathcal{X}^{\mathcal{A}}|}$. The entries are given by $G_{\indmap{}(\svbx{})\,\indmap{}(\svbx{}')} = p(\svbx{}\,|\,\svbx{}') \in [0,1]$, and encode the probability of the state transitions from $\svbx{}'$ to $\svbx{}$ in a single algorithmic step. These probabilities aggregate all microscopic realizations consistent with the corresponding coarse-grained update. Even when $\mathcal{A}$ is deterministic at the microscopic level, $\mathbf{G}_{\mathcal{A}}$ can be stochastic due to coarse-graining, where multiple microscopic configurations correspond to the same algorithmic state. For inherently stochastic algorithms, the entries additionally capture intrinsic algorithmic randomness.

Let $\px{i}$ be the probability distribution over the coarse-grained state space $\mathcal{X^A}$ at step $i$. The computational model corresponding to the algorithm can be formalized as a \emph{periodic machine}~\cite{wolpert24_cct2algo}, a discrete-time system in which the same transition matrix ${\bf G}_{\mathcal{A}}$ governs the state evolution at each step. The algorithmic state distribution evolves as
\begin{equation}
    \pbx{i} = \gA{}^{i}\,\pbx{0}, \quad i \ge 1.
    \label{eq:2.10}
\end{equation}

Each application of $\gA{}$ in \eqref{eq:2.10} advances $\px{i}$ by one logical execution step of the algorithm. The algorithm progresses through a finite set of program-counter indices; iterative structures and control flow result in repeated visits to some states before reaching the halting instruction. Correspondingly, the state evolution of the periodic machine reaches a steady-state distribution, $\pbx{\rm T} = \gA{}^{\rm T}\,\pbx{0} = \pi_{\gA{}}$, where $\rm T$ is the minimum number of steps required for periodic closure. In principle, $\pi_{\gA{}}$ can be characterized as the eigenvector of $\gA{}^{\rm T}$ associated with eigenvalue $1$. However, direct computation is challenging due to the exponential size of the state space. The step count $\rm T$ is invariant under re-initialization with any valid initial distribution, reflecting the intrinsic periodicity of the system.

Since the same physical process implements $\gA{}$ at each step, the thermodynamically optimal prior $\qx{0}$ (cf.~Definition~\ref{def:3}) is identical across steps. We therefore next define the algorithmic mismatch cost, as also introduced in~\cite{wolpert24_DTMC, wolpert24_cct2algo, wolpert25_RASP}.
\begin{definition}[Algorithmic mismatch cost~\cite{wolpert24_DTMC}]
    Let $\mathbf{G}_\mathcal{A}$ be the stochastic transition matrix of an algorithm 
    $\mathcal{A}$ halting in $\mathrm{T}$ steps, $\qx{0}$ the associated 
    prior distribution, and $\px{0}$ the actual initial distribution over $\mathcal{X}^\mathcal{A}$. 
    The step-wise mismatch cost incurred at step $i \in [\mathrm{T}]$ is defined as
    \begin{equation}
        \pmmc{\pbx{0}}  
        = \DKL{\gA{}^{i-1} \px{0}}{\qx{0}}
          - \DKL{\gA{}^{i} \px{0}}{\gA{} \qx{0}}.
        \label{eq:2.11}
    \end{equation}
    and the total algorithmic mismatch cost is
    \begin{equation}
        \mmc{\px{0}} = \sum\nolimits_{i=1}^{\mathrm{T}} \pmmc{\pbx{0}}.
        \label{eq:2.12}
    \end{equation}
\end{definition}

Even when $\px{0} = \qx{0}$, the distribution typically diverges from the prior after the first update unless $\gA{}$ is a permutation or otherwise degenerate. Mismatch cost, therefore, quantifies the unavoidable dissipation generated by repeated execution of an algorithmic update.

%%%%%%%%%%%%%%%%%%%%%%%%%%%%%%%%%%%%%%%%%%%%%%%%%%%%%%%%%%%%%%%%%%%%%%%%%%%%%%%%%%%%%%%%%%%%%%%%
\section{Problem setup}
\label{sec:framework}
\begin{figure}[!ht]
    \centering
    \includegraphics[width=0.9\linewidth]{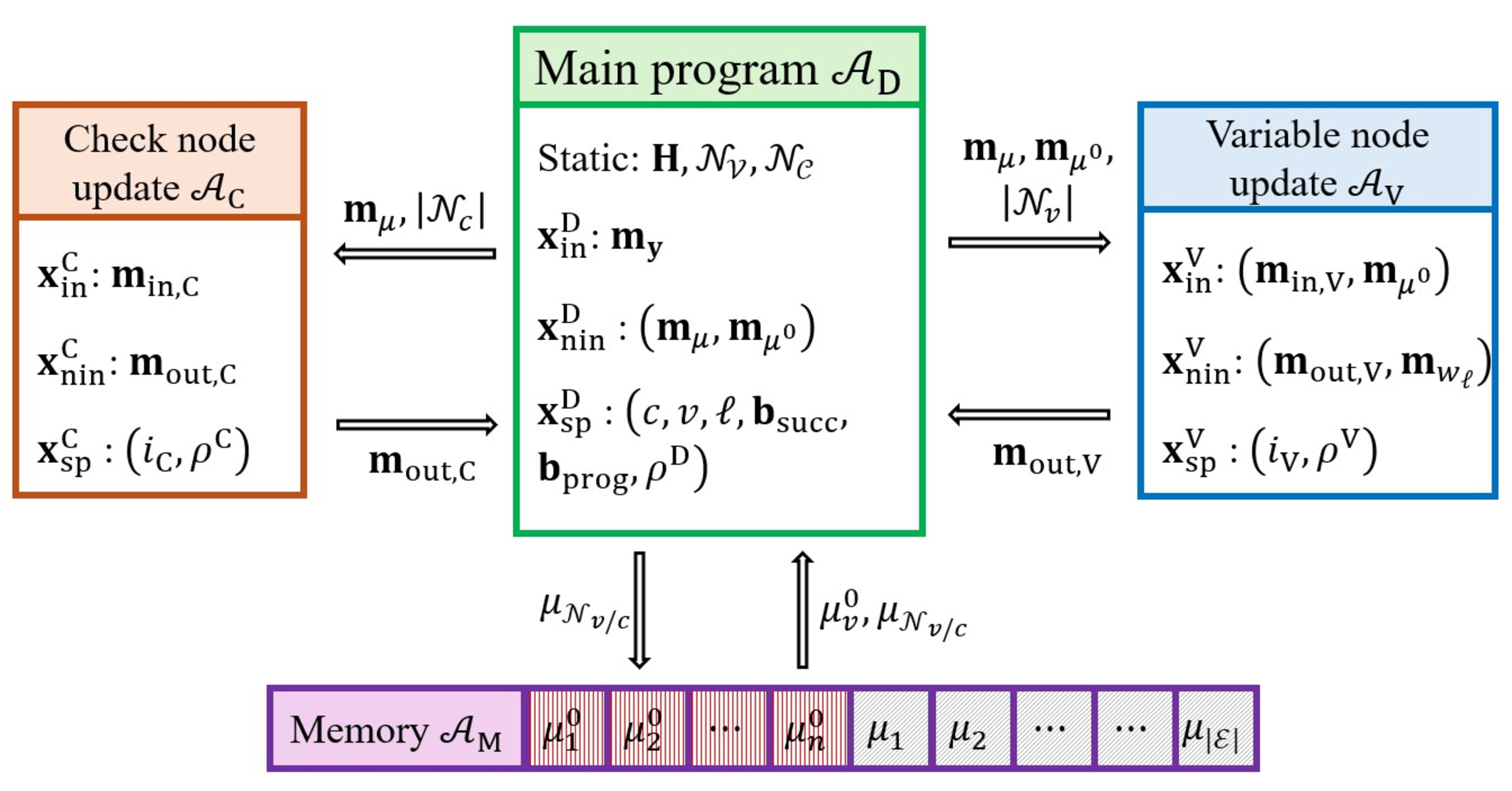}
    \caption{Overall system model of a message-passing decoder, with relevant registers. The main program $\A{D}$ interacts with subroutines $\A{C}$ and $\A{V}$. Memory $\A{M}$ holds channel observations in the first $n$ locations and messages exchanged along the Tanner graph in the next $|\mathcal{E}|$ locations.}
    \label{fig:1}
\end{figure}

In this section, we present our framework for analyzing the mismatch cost of message-passing decoders. Consider a linear code defined by a parity-check matrix $\mathbf{H}\in \mathbb{F}_2^{m\times n}$, with Tanner graph $G=(\mathcal{V},\mathcal{C},\mathcal{E})$. The decoder is modeled as three interacting periodic machines. The main decoder machine, $\A{D}$, orchestrates the overall decoding process. In addition, two devices, $\A{C}$ and $\A{V}$, implement the check-node and variable-node update subroutines, respectively. Each subroutine executes its corresponding update rules uniformly across all nodes of its type, as specified by the underlying message-passing algorithm. To support these computations, the messages along edges and channel observations are stored in an external memory subsystem $\A{M}$, a contiguous array of length $n + |\mathcal{E}|$ accessed by $\A{D}$ during each subroutine call. Read and write operations on $\A{M}$ are treated as simple copying operations, analogous to idealized channel communication~\cite{wolpert24_comchannel}. Fig.~\ref{fig:1} provides a high-level illustration of the system architecture.

We adopt a call-by-value model~\cite{wolpert24_cct2algo}, in which $\A{D}$ passes copies of message values to its subroutines, ensuring that local computations do not modify the state of $\A{D}$. Each device is implemented using the Random Access Stored Program (RASP) architecture~\cite{rasp}, a von Neumann–style architecture in which registers store both data and instructions. Instructions are executed sequentially, with conditional branches governing the flow, enabling deterministic, \emph{stepwise} computation~\cite{wolpert24_cct2algo}. Registers may store arbitrary symbols instead of binary values, enabling the architecture to represent computations directly on the symbols. This abstraction provides a fine-grained representation of state transitions, capturing the evolution of both messages and internal state for $\A{D}$ and its subroutines. Hereafter, we use $\A{A}$ for ${\rm A} \in \{\rm D, V, C\}$ to refer to both the algorithm and its corresponding periodic device, as appropriate.

\subsection{State space design for main program $\A{D}$}
\label{subsec:3.A}
We characterize the state space of the main program $\A{D}$ for iterative decoding. For a code defined by $G=(\mathcal{V},\mathcal{C},\mathcal{E})$, the sets-of-sets $\mathcal{N_V} = \{\mathcal{N}_v : v \in [n]\}$ and $\mathcal{N_C} = \{\mathcal{N}_c : c \in [m]\}$ capture the code’s structure. Although these sets are required for the execution of $\A{D}$, they are fixed parameters and therefore not part of the state space. 

The program maintains a collection of registers, including data and control variables such as flags, loop counters, and the program counter. These are updated throughout the iterative decoding process. Table~\ref{tab:mainprog_state} lists these registers, their sizes, their alphabets, and their roles in the decoding process. These registers collectively define the (random) state of $\A{D}$ at computational step $i$ as $\xAn{D}{i} = (\mr{y}, \mr{\mu}, \mr{\mu^0}, c, v, \ell, \bsp, \bss, \pc{D}) \in \XA{D}$, where $\XA{D} \subseteq \mathcal{O}^n \times \mathcal{M}^{d_{\max}} \times \mathcal{O} \times {[m]}_0 \times {[n]}_0 \times {[\ell_{\max}]}_0 \times \{0,1\}^2 \times {[\,\pcc{D}\,]}_0$ is the reachable state space. 
\begin{table}[H]
    \centering
    \caption{Registers defining the state of the main decoder program $\A{D}$.}
    \label{tab:mainprog_state}
    \begin{tabular}{p{5cm} >{\centering\arraybackslash}m{1.7cm} >{\centering\arraybackslash}m{1.7cm} >{\centering\arraybackslash}m{3cm} >{\centering\arraybackslash}m{2cm}}
    \toprule
    \textbf{State variable} & \textbf{Symbol} & \textbf{Size} & \textbf{Values} & \textbf{Type}\\
    \midrule
    Codeword reg. & $\mr{y}$ & $n$ & $\mathcal{O}^n$ & Input\\ \midrule
    Edge-message reg. & $\mr{\mu}$ & $d_{\max}$ & $\mathcal{M}^{d_{\max}}$ & Non-input\\ \midrule
    Variable observation reg. & $\mr{\mu^0}$ & $1$ & $\mathcal{O}$ & Non-input\\ \midrule
    Check-node loop counter & $c$ & $1$ & ${[m]}_0$ & Special\\ \midrule
    Variable-node loop counter & $v$ & $1$ & ${[n]}_0$ & Special\\ \midrule
    Message-pass loop counter & $\ell$ & $1$ & ${[\ell_{\max}]}_0$ & Special\\ \midrule
    Progress flag & $\bss$ & $1$ & $\{0,1\}$ & Special\\ \midrule
    Success flag & $\bsp$ & $1$ & $\{0,1\}$ & Special\\ \midrule
    Program counter & $\pc{D}$ & $1$ & ${[\,\pcc{D}\,]}_0$ & Special\\
    \bottomrule
    \end{tabular}
\end{table}

Under the RASP framework, each register can take values from a specified domain of symbols, enabling $\A{D}$ to accommodate diverse decoding algorithms and channel models. The \texttt{Size} column in Table~\ref{tab:mainprog_state} specifies the number of memory cells in each register, while the \texttt{Values} column indicates the set of possible overall register values. Specifically, the registers $\mr{y}$ and $\mr{\mu^0}$ consist of cells holding symbols from the observation alphabet $\mathcal{O}$, such that the overall register values belong to $\mathcal{O}^n$. In contrast, the register $\mr{\mu}$ comprises $d_{\max}$ cells holding symbols from the message alphabet $\mathcal{M}$, which depends on the decoding algorithm and the channel model. For instance, in the Gallager-A algorithm~\cite{richardson2008modern} or sum-product decoding over the binary-erasure channel (BEC), $\mathcal{O},\mathcal{M} = \{-1,0,1\}$. In contrast, sum-product decoding with log-likelihood representation employs messages in $\mathcal{M} \subset \mathbb{R}$, quantized in $b$-bit fixed-point format, resulting in $\mr{\mu}$ holding $d_{\max}$ quantized real values. Table~\ref{tab:mainprog_state} defines the general structure of the decoder state independently of these specific choices.

During decoding, $\A{D}$ reads edge messages and channel observations from memory $\A{M}$ into the registers $\mr{\mu}$ and $\mr{\mu^0}$. It updates these messages using the subroutines $\A{C}$ and $\A{V}$ and writes the results back to memory. The codeword estimate stored in $\mr{y}$ is then updated accordingly. Throughout this process, the program counter $\pc{D}$ tracks the current instruction among $\pcc{D}$ machine-level instructions stored in a linear instruction memory. Each instruction corresponds to an operation of the decoding algorithm. These operations include invoking subroutines, updating registers such as $\mr{y}$, and managing loop counters and message updates. Therefore, $\pcc{D}$ represents the instruction memory footprint required to implement the algorithm. Execution proceeds over all nodes and iterations until convergence or the maximum number of iterations $\ell_{\max}$ is reached. At that point, $\pc{D}$ reaches the final instruction, and the decoder halts.

Each computational step of $\A{D}$ updates the contents of its registers, thereby inducing a transition in the state space $\XA{D}$. This execution defines a state evolution over $\XA{D}$, which is captured by the state transition matrix $\gA{D} = [\,g_{\scriptscriptstyle dd'}\,] \in \mathbb{R}^{|\XA{D}| \times |\XA{D}|}$, where ${g}_{\scriptscriptstyle dd'} \in [0,1]$ denotes the probability of transitioning from $\iindmap{D}(d')$ to $\iindmap{D}(d)$. The matrix $\gA{D}$ is constructed by enumerating all states in $\XA{D}$ and recording the corresponding transition probabilities. Following \eqref{eq:2.10}, the distribution at computational step $i \in [\,{\rm T_D}\,]$ starting from the initial distribution $\pbxa{0}{D}$ is $\pbxa{i}{D} = \gA{D}^{i} \pbxa{0}{D}$.
As the algorithm halts after ${\rm T_D}$ computational steps, this distribution achieves the steady-state distribution $\pi_{\gA{D}}$, which is the distribution at the halting step, i.e., $\pi_{\gA{D}} = \pbxa{\rm T_D}{D}$.

\subsection{Initial distribution and mismatch cost of main program $\A{D}$}
\label{subsec:3.B}
In this section, we define the initial state distribution $p_{\xAn{D}{0}}$ of the main decoder program $\A{D}$ over its state space $\XA{D}$. As specified in Table~\ref{tab:mainprog_state}, the state of $\A{D}$ is the joint realization of input, internal (non-input), and special variables (e.g., program or loop counters) governing execution flow, with respective state spaces $\XA{D}_{\rm in}$, $\XA{D}_{\rm nin}$, and $\XA{D}_{\rm sp}$. Let $\rndvin{D}$, $\rndvint{D}$, and $\rndvsp{D}$ denote the corresponding random states, with realizations $\smpvin{D}$, $\smpvint{D}$, and $\smpvsp{D}$, respectively. The overall state is therefore $\xA{D} = (\rndvin{D}, \rndvint{D}, \rndvsp{D}) \in \XA{D}$, where $\XA{D} = \XA{D}_{\rm in} \times \XA{D}_{\rm nin} \times \XA{D}_{\rm sp}$.

The input state $\rndvin{D}$ of $\A{D}$ is the channel output $\mathbf{y} \in \mathcal{O}^n$. Its distribution $\pin{D}$ is induced by the message prior, the encoding map, and the channel. Let $\rvbu{} \in \mathbb{F}_2^k$ be a random message vector, where $k = n - \mathrm{rank}({\bf H})$. We assume $\rvbu{}$ has i.i.d. (but biased) components $\rvu{i} \sim \mathrm{Ber}(\alpha)$ where $\Pr(\rvu{i}=0)=\alpha$ denotes the \emph{message prior}. For any realization $\svbu{}$, let $\wt{\svbu{}}$ denote its Hamming weight. Then, $p_{\rvbu{}}(\svbu{}) = \alpha^{k-\wt{\svbu{}}}\cdot (1 - \alpha)^{\wt{\svbu{}}}$.
Under the generator matrix $\mathbf{G}_{\rm E}$ -- distinct from state-transition matrix $\gA{D}$ -- the transmitted codeword is $\rvbxu = \rvbu{}\mathbf{G}_{\rm E}$. Since encoding is deterministic, $p_{\rvbxu}(\svbxu) = p_{\rvbu{}}(\svbu{})$. With $p_{{\bf y} \,|\, \rvbxu}(\bm{y} \,|\, \svbxu)$ denoting the conditional probability of observing channel output $\bf y$ given $\rvbxu$, the resulting channel-output distribution is 
\begin{equation}
    p_{\bf y}(\bm{y}) = \sum_{\svbxu\in \mathbb{F}_2^{n}}\,\sum_{\svbu{} \in \mathbb{F}_2^{k}}
    p_{\bf y|\rvbxu}(\bm{y} \,|\, \svbxu)\delta(\svbxu, \svbu{} \mathbf{G}_{\rm E})
    p_{\rvbu{}}(\svbu{}).
    \label{eq:3.8}
\end{equation}
Since the decoder input state equals the channel output, we set $\rndvin{D} = \mathbf{y}$, so $\pin{D}(\smpvin{D}) = p_{\mathbf{y}}(\smpvin{D})$.

The special state variables (e.g., program counters) of $\A{D}$ are deterministically reset to a fixed initial state $\smpxA{D}_{\rm sp,0}$ at the start of each execution, independent of the input. Accordingly,
\begin{equation}
    p_{\rndvsp{D}}(\smpvsp{D}) = \delta(\smpvsp{D}, \smpxA{D}_{\rm sp,0}),
    \label{eq:3.4}
\end{equation}  
such that each $\smpvsp{D}$ is initialized to fixed state $\smpxA{D}_{\rm sp,0}$ (a system parameter) at the start of each run.

The internal (non-input) state variables are not reinitialized when a new input is processed. Instead, their initial distribution is the marginal of the joint state distribution at the end of the previous execution after $\rm T_D$ steps:
\begin{equation}
    p_{\rndvint{D}}(\smpvint{D})
    = \sum_{\smpvin{D}\in \XA{D}_{\rm in}}~\sum_{\smpvsp{D}\in \XA{D}_{\rm sp}} p_{\xAn{D}{\rm T_D}}\left(\smpvin{D}, \smpvint{D}, \smpvsp{D}\right),
    \label{eq:3.5}
\end{equation}
where the summation is over input and special variables of the state $\smpxA{D}=(\smpvin{D}, \smpvint{D}, \smpvsp{D}) \in \XA{D}=\XA{D}_{\rm in} \times \XA{D}_{\rm nin} \times \XA{D}_{\rm sp}$. This models a setting in which the internal state is carried forward across successive executions without reinitialization. Such retention of state is consistent with modern digital systems, where state variables -- implemented via flip-flop–based registers -- persist unless explicitly reset. Under this assumption, no implicit information erasure occurs, and hence no energy is expended on re-initialization, in accordance with the Landauer-Bennett principle \cite{bennett1982}. Consequently, re-initialization is confined to the input variables, which for $\A{D}$ correspond to the channel outputs. Since these inputs are supplied by an external system (the channel), the associated reinitialization cost is excluded from the algorithm's thermodynamic cost.

Using~\eqref{eq:3.8}, \eqref{eq:3.5}, and \eqref{eq:3.4}, the joint initial distribution at the start of the execution for $\A{D}$ is
\begin{equation}
    \vspace{.1cm}
    p_{\xAn{D}{0}}\left(\smpvin{D}, \smpvint{D}, \smpvsp{D}\,\right) = \pin{D}(\smpvin{D}) \, p_{\rndvint{D}}(\smpvint{D}) \, \delta(\smpvsp{D},  \smpxA{D}_{\rm sp, 0}),
    \label{eq:3.11}
\end{equation}
where ${\smpxA{D}=(\smpvin{D}, \smpvint{D}, \smpvsp{D}) \in \XA{D}=\XA{D}_{\rm in} \times \XA{D}_{\rm nin} \times \XA{D}_{\rm sp}}$.
The mismatch cost for $\A{D}$ for a prior $\qbxa{0}{D}$ over $\XA{D}$ is then
\begin{equation}
    \mmcA{D}{\pxa{0}{D}} = \sum\nolimits_{i=1}^{\rm T_D}\Big[\DKL{\gA{D}^{i-1}\pbxa{0}{D}}{\qbxa{0}{D}} - \DKL{\gA{D}^{i}\pbxa{0}{D}}{\gA{D}\qbxa{0}{D}}\Big],
    \label{eq:3.12}
\end{equation}
where $\rm T_D$ denotes the number of steps required for convergence to the steady-state distribution $\pi_{\gA{D}}$. Looking ahead to Sec.~\ref{subsec:4.A}, we note that $\rm T_D \leq 300$ in our implementation.

\subsection{State space design for the subroutines $\A{C}$ and $\A{V}$}
\label{subsec:3.C}
In this subsection, we characterize the state spaces of the subroutines $\A{C}$ and $\A{V}$. As in the main program $\A{D}$ (cf. Sec.~\ref{subsec:3.A}), each subroutine maintains a set of internal registers, including data registers and control variables such as loop counters and program counters. These registers define the local state of the computation. The constituent registers of $\A{C}$ and $\A{V}$ are listed in Tables~\ref{tab:chksubroutine_state}
, together with their alphabets, sizes, and roles.
\begin{table}[H]
    \centering
    \caption{State variables that defines the state of check-node subroutine $(\A{C})$ and variable-node subroutine $(\A{V})$}
    \label{tab:chksubroutine_state}
    \begin{tabular}{p{1.8cm} > {\raggedright\arraybackslash}m{5cm}  >{\centering\arraybackslash}m{1.2cm} >{\centering\arraybackslash}m{1.2cm} >{\centering\arraybackslash}m{2.7cm} >
    {\centering\arraybackslash}m{2cm}}
    \toprule
    \textbf{Subroutine} & \textbf{State variable} & \textbf{Symbol} & \textbf{Size} & \textbf{Values} & \textbf{Type}\\
    \midrule
    \multirow{4}{4em}{\rotatebox{90}{\parbox{3cm}{\centering Check-node ($\A{C}$)}}} & Incoming edge-message reg. & $\mr{in, \scrl{C}}$ & $d_{\max}$ & $\mathcal{M}^{d_{\max}}$ & Input\\ \noalign{\vskip 5pt} \cline{2-6} \noalign{\vskip 5pt}
    & Outgoing edge-message reg. & $\mr{out, \scrl{C}}$ & $d_{\max}$ & $\mathcal{M}^{d_{\max}}$ & Non-input\\ \noalign{\vskip 5pt} \cline{2-6} \noalign{\vskip 5pt}
    & Check-node loop counter & $i_{\scrl{\rm C}}$ & $1$ & ${[d_{\scrl{\mathcal{C}}}]}_0$ & Special\\ \noalign{\vskip 5pt} \cline{2-6} \noalign{\vskip 5pt}
    & Program counter & $\pc{C}$ & $1$ & ${[\,\pcc{C}\,]}_0$ & Special\\ \noalign{\vskip 5pt}
    \hlineB{3} \noalign{\vskip 5pt}
    \multirow{4}{6em}{\rotatebox{90}{\parbox{3cm}{\centering Variable-node ($\A{V}$)}}} & Incoming edge-message reg. & $\mr{in, \scrl{V}}$ & $d_{\max}$ & $\mathcal{M}^{d_{\max}}$ & Input\\ \noalign{\vskip 5pt} \cline{2-6} \noalign{\vskip 5pt}
    & Observation message reg. & $\mr{0}$ & $1$ & $\mathcal{O}$ & Input\\ \cline{2-6} \noalign{\vskip 5pt}
    & Outgoing edge-message reg. & $\mr{out,\scrl{V}}$ & $d_{\max}$ & $\mathcal{M}^{d_{\max}}$ & Non-input\\ \noalign{\vskip 5pt} \cline{2-6} \noalign{\vskip 5pt}
    & Weight reg. (optional) & $\mr{w_\ell}$ & $1$ & $\{w_1, \dots, w_{\ell_{\max}}\}$ & Non-input\\ \noalign{\vskip 5pt} \cline{2-6} \noalign{\vskip 5pt}
    & Variable-node loop counter & $i_{\scrl{\rm V}}$ & $1$ & ${[d_{\scrl{\mathcal{V}}}]}_0$ & Special\\ \noalign{\vskip 5pt} \cline{2-6} \noalign{\vskip 5pt}
    & Program counter & $\pc{V}$ & $1$ & ${[\,\pcc{V}\,]}_0$ & Special\\
    \bottomrule
    \end{tabular}
\end{table}

The (random) state of subroutine $\A{C}$ at step $i$ is $\xAn{C}{i} = (\mr{in, \scrl{C}}, \mr{out, \scrl{C}}, i_{\scrl{\rm C}}, \pc{C}) \in \XA{C}$, while that of $\A{V}$ is $\xAn{V}{i} = (\mr{in, \scrl{V}}, \mr{out, \scrl{V}}, \mr{0}, \mr{w_\ell}, i_{\scrl{\rm V}}, \pc{V}) \in \XA{V}$. The corresponding reachable state spaces are
$\XA{C} \subseteq \mathcal{M}^{d_{\max}} \times \mathcal{M}^{d_{\max}} \times {[d_{\scrl{\mathcal{C}}}]}_0 \times {[\,\pcc{C}\,]}_0$ and
$\XA{V} \subseteq \mathcal{M}^{d_{\max}} \times \mathcal{M}^{d_{\max}} \times \mathcal{O} \times \{w_1, \dots, w_{\ell_{\max}}\} \times {[d_{\scrl{\mathcal{V}}}]}_0 \times {[\,\pcc{V}\,]}_0$, where $\pcc{C}$ and $\pcc{V}$ denote the number of machine-level instructions stored in the linear instruction memory of $\A{C}$ and $\A{V}$, respectively. 

In the following, we use $\rm A \in \{V, C\}$ to denote a generic subroutine, which enables a unified description. The domains of the input and output registers $\mr{in, \scrl{A}}$ and $\mr{out, \scrl{A}}$ depend on the message-passing algorithm and the channel model as defined in Tables~\ref{tab:chksubroutine_state}. The treatment of non-input and special state variables follows directly from Sec.~\ref{subsec:3.A} and is not repeated here.

\begin{figure}[!ht]
    \centering 
    \subfloat[System-level interaction for check-node computation.]{
    \includegraphics[width=0.52\textwidth]{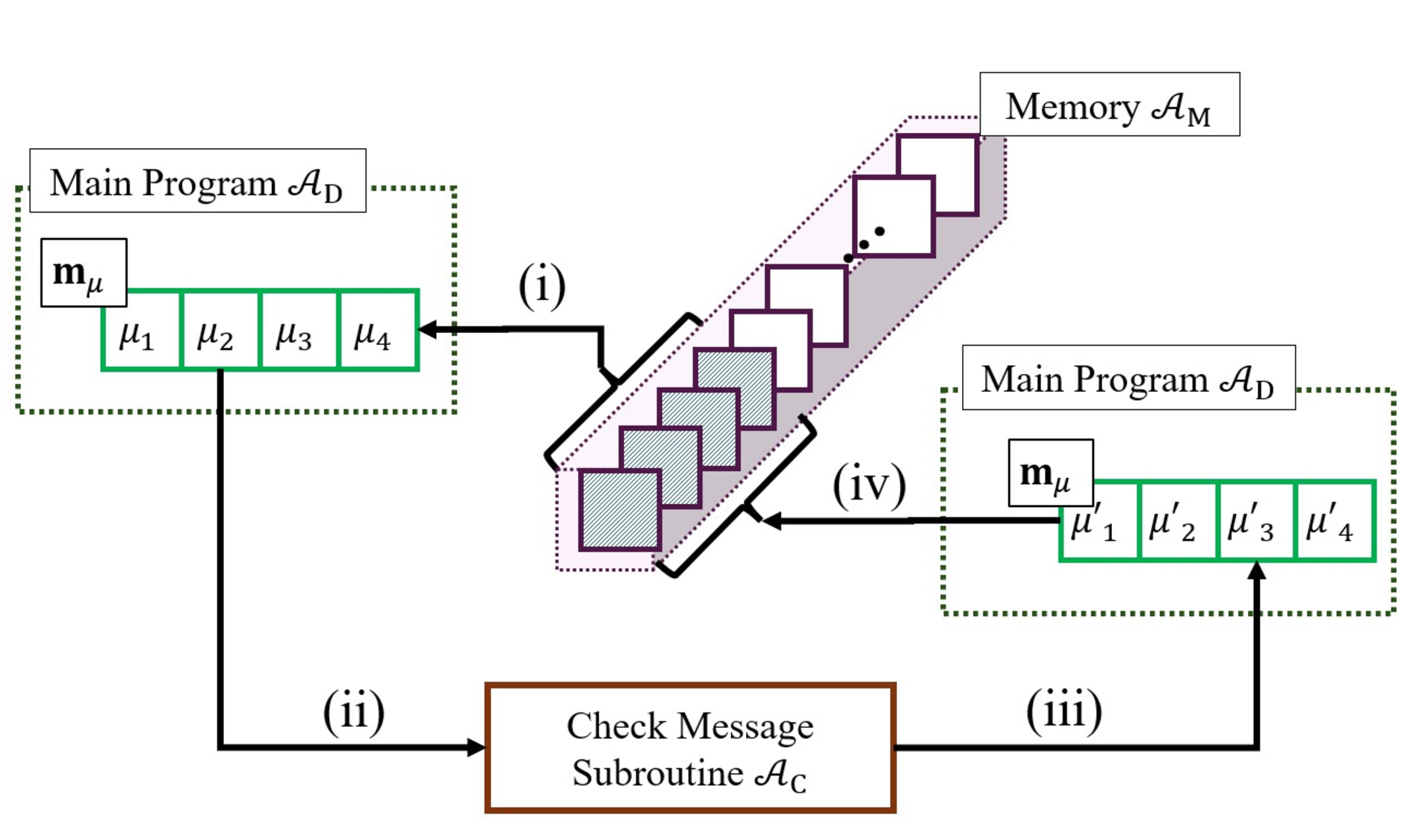}
    \label{subfig:2.1}
    }
    \subfloat[State-level interaction for check-node computation.]{
    \includegraphics[width=0.48\textwidth]{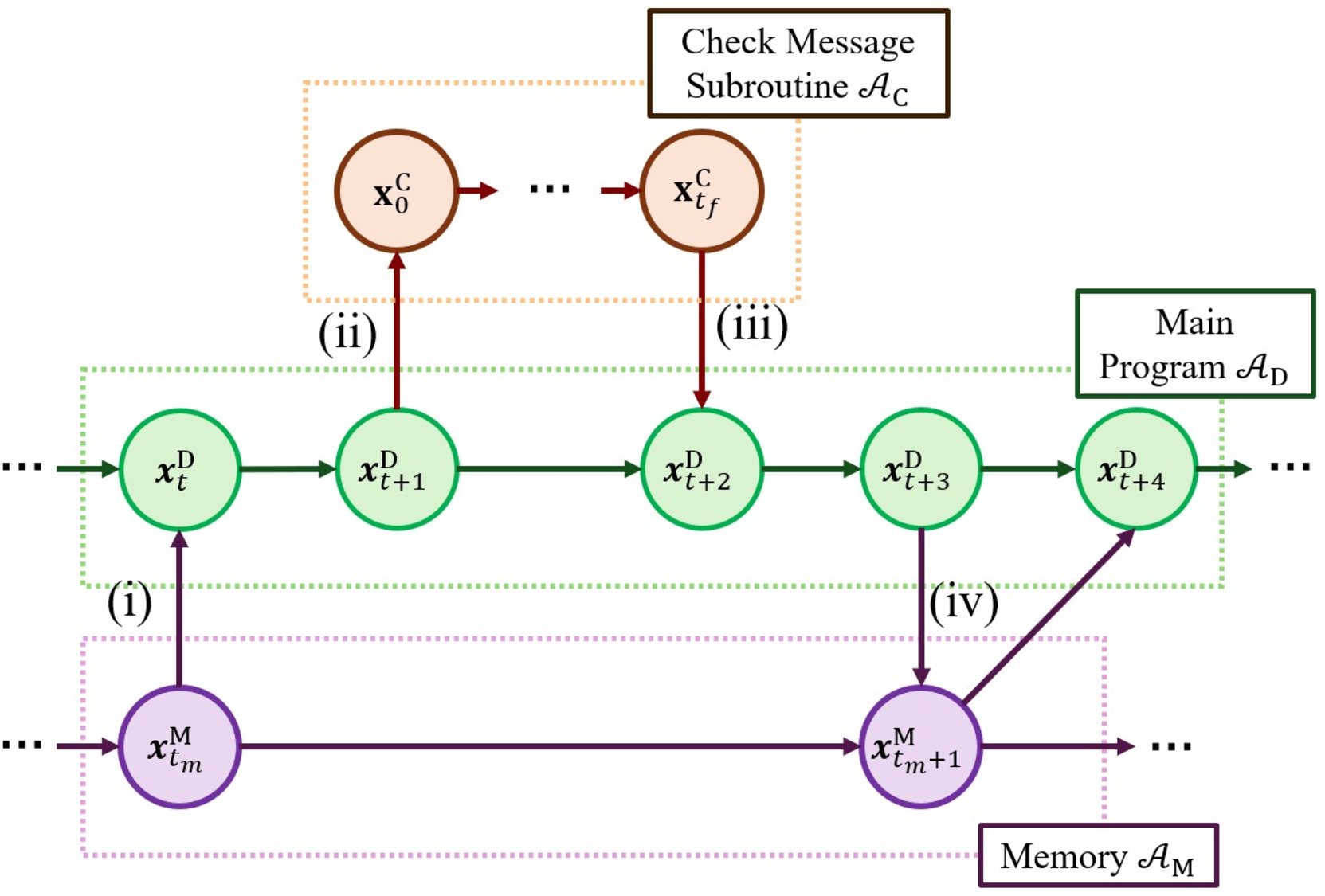}
    \label{subfig:2.2}
    }
    \caption{Interactions between decoder $\A{D}$, memory $\A{M}$, and check-node subroutine $\A{C}$: (i) $\A{D}$ retrieves messages from $\A{M}$ onto $\mr{\mu}$, (ii) forwards $\mr{\mu}$ to $\mr{in, \scrl{C}}$ of $\A{C}$, (iii) $\mr{\mu}$ is updated using $\mr{out, \scrl{C}}$ from $\A{C}$, and (iv) $\A{D}$ writes $\mr{\mu}$ back to $\A{M}$.}
    \label{fig:2}
\end{figure}

\begin{figure}[!ht]
    \centering 
    \subfloat[System-level interaction for variable-node computation.]{
    \includegraphics[width=0.5\textwidth]{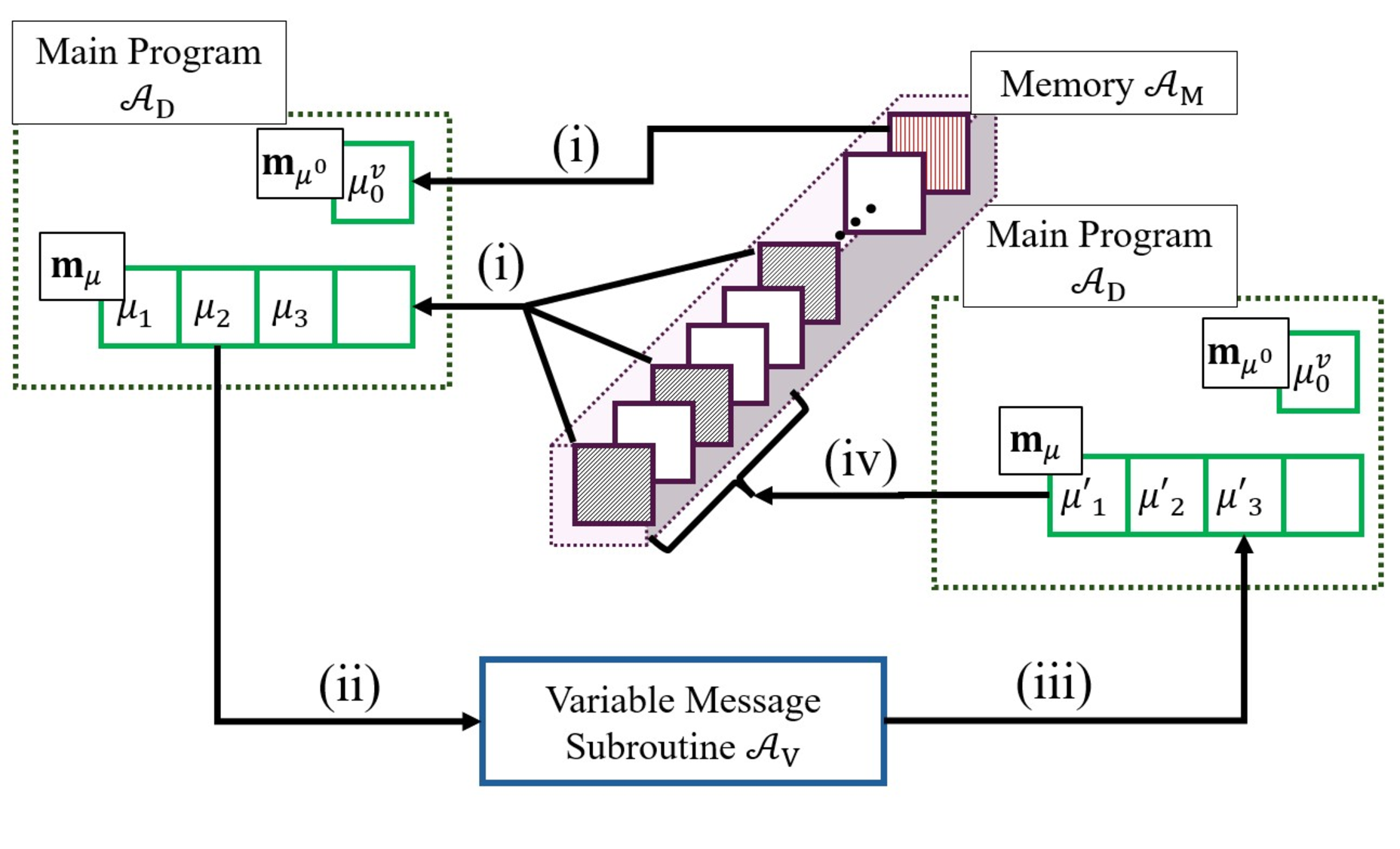}
    \label{subfig:3.1}
    }
    \subfloat[State-level interaction for variable-node computation.]{
    \includegraphics[width=0.5\textwidth]{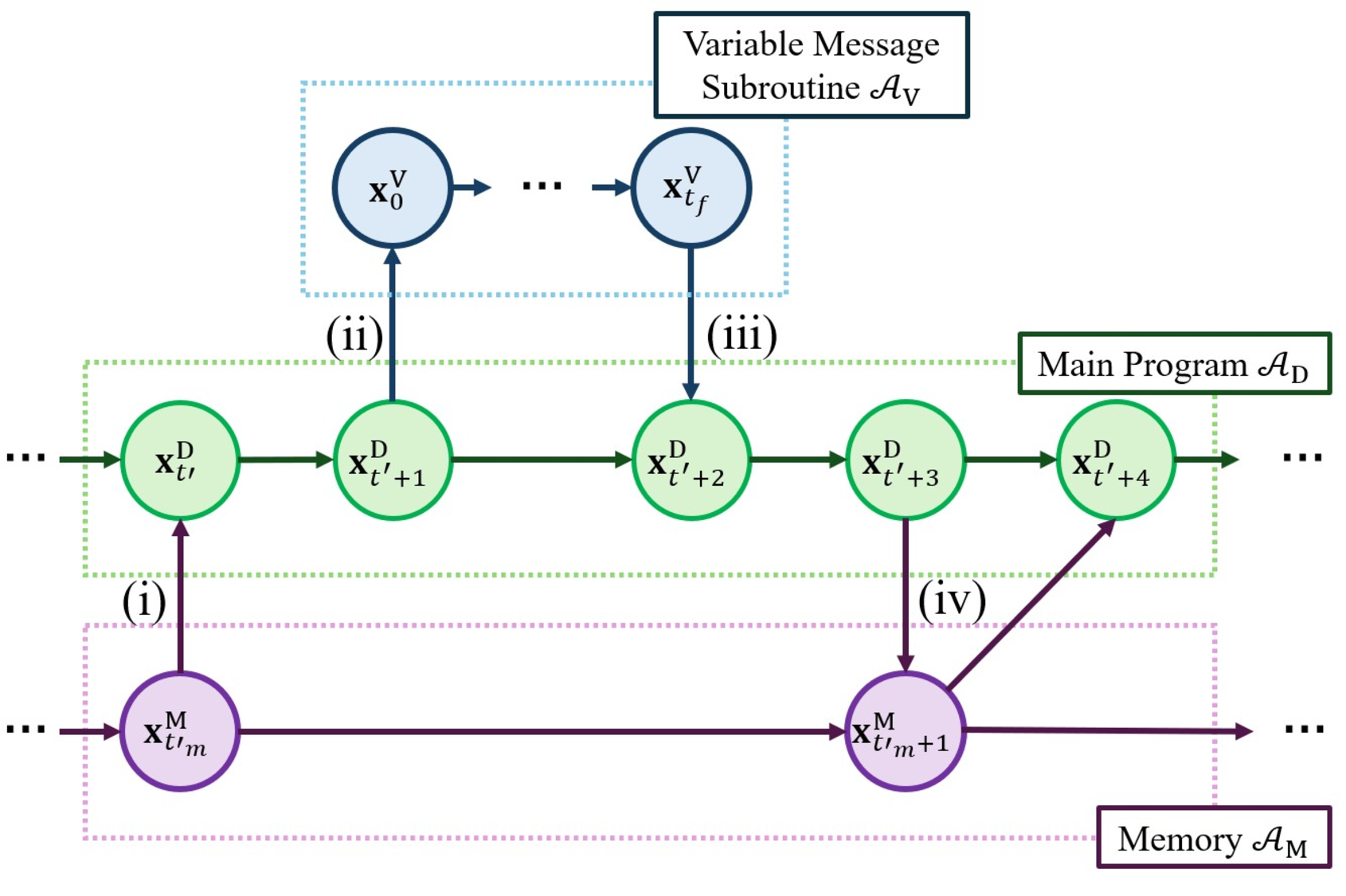}
    \label{subfig:3.2}
    }
    \caption{Interactions between decoder $\A{D}$, memory $\A{M}$, and variable-node subroutine $\A{V}$: (i) $\A{D}$ retrieves messages from $\A{M}$ onto $\mr{\mu}$ and $\mr{\mu^0}$, (ii) forwards $\mr{\mu}$ and $\mr{\mu^0}$ to $\mr{in, \scrl{V}}$ and $\mr{0}$ of $\A{V}$ respectively, (iii) $\mr{\mu}$ is updated using $\mr{out, \scrl{V}}$ from $\A{V}$, and (iv) $\A{D}$ writes $\mr{\mu}$ back to $\A{M}$.}
    \label{fig:3}
\end{figure}

Both subroutines implement local message updates over node neighborhoods. The input registers collect incoming messages from neighboring nodes, together with the observation $\mr{0}$ in the case of $\A{V}$. The subroutine then executes a fixed sequence of $\pcc{A}$ machine-level instructions to produce updated outgoing messages $\mr{out, \scrl{A}}$. These outputs are returned to $\A{D}$ and written to memory $\A{M}$. The interaction patterns with $\A{D}$ and $\A{M}$ are illustrated in Fig.~\ref{fig:2} and Fig.~\ref{fig:3}, and are structurally identical for both subroutines.

The execution of each subroutine induces a state evolution over $\XA{A}$. This evolution is characterized by a state transition matrix $\gA{A} = [g_{aa'}] \in \mathbb{R}^{|\XA{A}| \times |\XA{A}|}$, where $g_{aa'} \in [0, 1]$ denotes the probability of transitioning from $\iindmap{A}(a')$ to $\iindmap{A}(a)$. The construction of $\gA{A}$ follows the same procedure as for $\gA{D}$, by enumerating all states in $\XA{A}$ and recording the corresponding transition probabilities. The distribution at computational step $i \in [\,{\rm T_A}\,]$ starting from the initial distribution $\pbxa{0}{A}$ is $\pbxa{i}{A} = \gA{A}^{i} \pbxa{0}{A}$. 
As each subroutine halts after ${\rm T_A}$ steps, this distribution converges to the steady-state distribution $\pi_{\gA{A}}$, which is the distribution at the halting step for each subroutine, i.e., $\pi_{\gA{A}} = \pbxa{\rm T_A}{A}$.

\subsection{Initial distribution and mismatch cost of subroutines $\A{C}$ and $\A{V}$}
\label{subsec:3.D}
For the subroutines $\A{V}$ and $\A{C}$, we assume a uniform distribution over their input spaces, reflecting that all incoming message configurations along Tanner-graph edges are admissible:
\begin{equation}
    p_{\rndvin{A}}(\smpvin{A}) = \frac{1}{\vert\XA{A}_{\rm in}\vert}, \qquad \forall \smpvin{A} \in \XA{A}_{\rm in},
    \label{eq:3.13}
\end{equation}
for $\rm A \in \{V, C\}$, where $\rndvin{V} = (\mr{in, \scrl{V}},\mr{0})$ and $\rndvin{C} = \mr{in, \scrl{C}}$.

Similar to the discussion in Sec.~\ref{subsec:3.B}, special state variables are deterministically reset to a fixed initial configuration $\smpxA{A}_{\rm sp,0}$ at the start of each execution, independent of the input (cf.~\eqref{eq:3.4}). Furthermore, non-input internal variables are not reinitialized across executions; their initial distribution is given by the marginal of the terminal distribution at step $\rm T_A$ (cf.~\eqref{eq:3.5}). Hence,
\begin{equation}
    p_{\rndvsp{A}}(\smpvsp{A}) = \delta(\rndvsp{A}, \smpxA{A}_{\rm sp,0}), \quad 
    p_{\rndvint{A}}(\smpvint{A}) = \sum\nolimits_{\sim \{\smpvint{A}\}} p_{\xAn{A}{\rm T_A}}(\smpxA{A}),
    \label{eq:3.14}
\end{equation}
for $\rm A \in \{V, C\}$.
Combining \eqref{eq:3.13} and \eqref{eq:3.14}, the joint initial distribution for each subroutine $\rm A \in \{V, C\}$ can be defined as
\begin{equation}
    p_{\xAn{A}{0}}(\smpxA{A}) = \frac{1}{\vert\XA{A}_{\rm in}\vert} 
    \left(\sum\nolimits_{\sim\{\smpvint{A}\}} p_{\xAn{A}{\scriptscriptstyle {\rm T_{A}}}}(\smpxA{A})\right)
    \delta(\rndvsp{A}, \smpxA{A}_{\rm sp, 0}),
    \label{eq:3.16}
\end{equation}
where $\rndvint{V} = \mr{out, \scrl{V}}$, $\rndvint{C} = \mr{out, \scrl{C}}$, and $\rm T_A$ denotes the number of steps required for convergence to the steady-state distribution $\pi_{\gA{A}}$. In our implementation, $\rm T_V \leq 8$ and $\rm T_C \leq 10$.
The mismatch cost of each subroutine $\rm A \in \{V, C\}$ with respect to its prior $\qbxa{0}{A}$ is therefore written as
\begin{equation}
    \mmcA{A}{\pbxa{0}{A}} = \sum\nolimits_{i=1}^{\rm T_{A}}\Big[\DKL{\gA{A}^{i-1}\pbxa{0}{A}}{\qbxa{0}{A}} - \DKL{\gA{A}^{i}\pbxa{0}{A}}{\gA{A}\qbxa{0}{A}}\Big].
    \label{eq:3.18}
\end{equation}

\subsection{Total mismatch cost of the message-passing decoder}
\label{subsec:3.E}
Following \cite{wolpert25_RASP}, the total mismatch cost of the message-passing decoder, accounting for the interactions between $\A{D}$, $\A{V}$, and $\A{C}$, is written as
\begin{equation}
    \mmcA{Total}{\pbxa{0}{D}, \pbxa{0}{V}, \pbxa{0}{C}} 
    = \mmcA{D}{\pbxa{0}{D}} + \ell_{\max}\left[n\,\mmcA{V}{\pbxa{0}{V}} + m\,\mmcA{C}{\pbxa{0}{C}}\right].
    \label{eq:3.19}
\end{equation}
The factors $n$ and $m$ account for the number of variable- and check-node updates per message-passing iteration. For each iteration, the subroutines contribute a mismatch cost of $n\,\mmcA{V}{\pbxa{0}{V}} + m\,\mmcA{C}{\pbxa{0}{C}}$. Since we assume no early termination, the decoder runs for $\ell_{\max}$ iterations, so the mismatch contribution of subroutines scales linearly with $\ell_{\max}$.

\subsection{Memory as an external subsystem $\A{M}$}
\label{subsec:3.F}
The memory subsystem $\A{M}$ is modeled as a contiguous array $\xA{M} \in \XA{M} \subseteq \mathcal{O}^n \times \mathcal{M}^{|\mathcal{E}|}$ of length $n + |\mathcal{E}|$, with the first $n$ entries storing channel observations and the remaining $|\mathcal{E}|$ holding messages along Tanner-graph edges. $\A{M}$ acts as a shared external resource accessed by $\A{D}$ during message updates (Fig.~\ref{fig:1}), with read/write operations treated as simple copying, analogous to idealized channel communication \cite{wolpert24_comchannel}. While memory access can dominate latency and energy in real systems, here we treat $\A{M}$ as an external subsystem rather than as part of the main program whose thermodynamic cost is analyzed. Incorporating memory as an active thermodynamic component, capturing the cost of data movement, is beyond the scope of this work but could be addressed in future extensions.

\subsection{Correctable vs. non-correctable erasure patterns and their initial distributions} 
\label{subsec:3.G}
The input distribution $\pbxa{0}{D}$ in \eqref{eq:3.11} is defined over all channel outputs and corresponding erasure patterns. In this subsection, we refine this distribution by conditioning on whether a given erasure pattern is correctable or non-correctable under the message-passing decoder within $\ell_{\max}$ message-passing iterations.
Importantly, we assume that the decoder hardware remains unchanged. Consequently, the state space $\XA{D}$, state transition matrix $\gA{D}$, and prior distribution $\qbxa{0}{D}$ are fixed. The same main program $\A{D}$ is executed, while the input distribution is restricted to either the correctable or non-correctable subset. This is analogous to operating a general-purpose processor on different input ensembles.

Let $\XAE{D}{E} \subseteq \XAin{D}$ denote the set of correctable erasure patterns and $\XAE{D}{H} \subseteq \XAin{D}$ the set of non-correctable ones. For example, in a $[7,4]$ Hamming code over a binary erasure channel, all weight-$2$ erasure patterns are correctable under sum-product decoding, while higher-weight patterns (with more than $4$ erasures) may be non-correctable and belong to a stopping set. Conditioning $p_{\rndvin{D}}$ on these subsets yields
\begin{equation}
    p_{\rndvin{D}\,\mid\,\rndvin{D} \in \XAE{D}{\chi}}(\smpvin{D}) = p_{\XAE{D}{\chi}}(\smpvin{D}) = \frac{p_{\rndvin{D}}(\smpvin{D}) \, \ind{\smpvin{D} \in \XAE{D}{\chi}}}{\sum\nolimits_{\smpvin{D} \in \XAE{D}{\chi}} p_{\rndvin{D}}(\smpvin{D})\ind{\smpvin{D} \in \XAE{D}{\cdot}}}, \quad \chi \in \{\rm E, H\}.
    \label{eq:3.20}
\end{equation}

These conditional input distributions are used to initialize the input state variables of the decoder in \eqref{eq:3.11}. With $\smpxA{D} = (\smpvin{D}, \smpvint{D}, \smpvsp{D}) \in \XA{D}$, the initial state distribution for correctable and non-correctable patterns takes the form
\begin{equation}
    p_{\xAn{D}{0}}(\smpxA{D}) = p_{\XAE{D}{\chi}}(\smpvin{D}) \, \left(\sum\nolimits_{\smpvin{D} \in \XAE{D}{\chi}}\sum\nolimits_{\smpvsp{D} \in \XA{D}_{\rm sp}} p_{\xAn{D}{\rm T_D}}\left(\smpvin{D}, \smpvint{D}, \smpvsp{D}\right)\right) \, \delta(\smpvsp{D}, \smpxA{D}_{\rm sp, 0}).
    \label{eq:3.21}
\end{equation}
Together with the prior $\qbxa{0}{D}$, these distributions allow \eqref{eq:3.12} to quantify the mismatch costs separately for each error category.  

%%%%%%%%%%%%%%%%%%%%%%%%%%%%%%%%%%%%%%%%%%%%%%%%%%%%%%%%%%%%%%%%%%%%%%%%%%%%%%%%%%%%%%%%%%%%%%%%
\section{Experimental Evaluations}
\label{sec:results}
We now present numerical results on mismatch costs for the message-passing decoder. Section~\ref{subsec:4.A} summarizes the experimental setup. In Sec.~\ref{subsec:4.B}, we evaluate mismatch costs under a \emph{uniform prior}, that corresponds to a physical implementation with approximately equal state-dependent entropy flows $f(\smpxA{A})$ into the reservoirs for all states $\smpxA{A} \in \XA{A}$, $\rm A \in \{\rm D, V, C\}$. This isolates mismatch cost from energetic biases across module states. Section~\ref{subsec:4.D} considers a \emph{non-uniform prior} $\qbtxa{0}{D}$ for the main decoder program $\A{D}$, induced by implementations in which $f(\smpxA{D})$ varies across $\smpxA{D} \in \XA{D}$. Consequently, states associated with larger entropy flows are assigned lower probabilities in $\qbtxa{0}{D}$. This shifts the combinations of message prior and erasure probability that minimize mismatch cost relative to the uniform case. Finally, Sec.~\ref{subsec:4.C} compares mismatch costs for correctable and non-correctable error patterns, illustrating how differences in their respective state-transition structures affect the overall cost.

\subsection{Experimental setup}
\label{subsec:4.A}
As described earlier, we illustrate the mismatch-cost framework for message-passing decoders using the $[7,4]$ Hamming code. Its small block length keeps the main program state space $\XA{D}$ tractable. This helps us to be able to examine key decoding mechanisms and their thermodynamic implications while initiating the development of methods that could scale to larger codes. Simulations are performed over a BEC with erasure probability $\epsilon \in [0,1]$. Binary information symbols $x \in \{0,1\}$ are mapped to the channel and message alphabets $\mathcal{O} = \mathcal{M} = \{-1,0,1\}$ via $x \mapsto (-1)^x$, with $0$ denoting an erasure. Channel observations and edge messages are stored in a contiguous array of $19$ cells: the first $7$ cells store channel outputs $\{\mu^0_v : v \in [7]\}$ and the remaining $12$ cells store edge messages $\{\mu_j : j \in [12]\}$ (cf.~Fig.~\ref{fig:1}). For the $[7,4]$ Hamming code, $d_{\max}=4$, and $\ell_{\max}=3$ suffices to ensure convergence.

The program counters $\rho^{\rm D}$, $\rho^{\rm C}$, and $\rho^{\rm V}$ index the $\pcc{D} = 26$, $\pcc{C} = 3$, and $\pcc{V} = 3$ unique machine-level instructions of $\A{D}$, $\A{C}$ and $\A{V}$ respectively. With this setup, the state space of $\A{D}$ is ${\XA{D} \subseteq \{-1, 0, 1\}^7 \times \{-1, 0, 1\}^4 \times \{-1, 0, 1\} \times {[3]}_0 \times {[7]}_0 \times {[3]}_0 \times \{0,1\}^2 \times {[26]}_0}$. 
As noted in Sec.~\ref{subsec:3.A}, the Cartesian state space of $\A{D}$ grows exponentially with $n$. For this small code with $n=7$, the Cartesian space has roughly $7.3$ billion states. Of these, only about $4.2$ million are reachable (i.e., states that are visited under the dynamics induced by $\gA{D}$), giving $|\XA{D}| \approx 4.2\,{\rm M}$. This is a central reason why we focus on a small code for our initial explorations.

The state transition matrix $\gA{D} \in \mathbb{R}^{4.2\mathrm{M} \times 4.2\mathrm{M}}$ is constructed as described in Sec.~\ref{subsec:3.B} by enumerating all combinations $(\mr{y}, \mr{\mu}, \mr{\mu^0}) \in \{-1,0,1\}^{12}$, with the special variables $\smpvsp{D}$ fixed to their predefined initial values, producing $3^{12}$ initial states. Each combination was simulated until termination to identify reachable states and transitions, which were then normalized column-wise to obtain $\gA{D}$. Even for the $[7,4]$ Hamming code, the reachable state space is sufficiently large to require careful memory management. To address this, $\gA{D}$ is partitioned into submatrices of size $\kappa = 3^{10}$, and the distributions are computed block-wise. Although this approach can, in principle, be extended to larger LDPC codes, the exponential growth of the decoding state space with block length makes it increasingly difficult to simulate the contribution of individual decoding trajectories to the overall thermodynamic cost.

The reachable state spaces for the check- and variable-update subroutines are considerably smaller under the simulation parameters described earlier, satisfying $|\XA{C}| < 10^{5}$ and $|\XA{V}| < 6.3 \times 10^{5}$. The corresponding transition matrices $\gA{C}$ and $\gA{V}$ can therefore be computed directly without block decomposition. Initial distributions $\pbxa{0}{D}$, $\pbxa{0}{V}$, and $\pbxa{0}{C}$ are defined according to \eqref{eq:3.11} and \eqref{eq:3.16}. Each periodic machine is assumed to interact with a single thermal reservoir ($R=1$) at temperature $T=300\,{\rm K}$, allowing mismatch cost to be expressed in Joules as $E(\pbxa{0}{A}) = k_{B}T\mmcA{A}{\pbxa{0}{A}}$. Unless stated otherwise, we assume the thermodynamically optimal prior $\qbxa{0}{A}$ to be uniform over $\XA{A}$ for each program ${\rm A}\in \{\rm D,V,C\}$. Equivalently, we assume the underlying implementation induces approximately uniform state-dependent entropy flows $f(\smpxA{A}) = \beta\Bar{Q}(\smpxA{A})$, such that all states are thermodynamically equal and the uniform distribution over $\XA{A}$ minimizes~\eqref{eq:2.6}. This allows the mismatch cost to reflect the effects of the initial distribution $\pbxa{0}{A}$ and decoder state transitions $\gA{A}$.

To provide deeper insight, and per Sec.~\ref{subsec:3.F}, we classify erasure patterns by whether or not they are correctable by message passing. Consider a received sequence $\mathbf{y}$ and let $\mathcal{I}_{\mathbf{y}} \subseteq \{1, \dots, 7\}$ denote the set of erased positions with $\mathcal{I}_{\mathbf{y}} \neq \emptyset$. The sets of correctable and non-correctable erasure patterns are defined as (``E'' for easy/correctable and ``H'' for hard/non-correctable) $\XA{D}_{\rm \mathcal{E}_E} = \left\{ \mathbf{y} \;\middle|\; |\mathcal{I}_{\mathbf{y}}| \le 2 \;\; \text{or} \;\; \big(|\mathcal{I}_{\mathbf{y}}| = 3 \;\text{and}\; \mathcal{I}_{\mathbf{y}} \notin \mathcal{S}\big) \right\}$ and $\XA{D}_{\rm \mathcal{E}_H} = \left\{ \mathbf{y} \;\middle|\; |\mathcal{I}_{\mathbf{y}}| \ge 4 \;\; \text{or} \;\; \big(|\mathcal{I}_{\mathbf{y}}| = 3 \;\text{and}\; \mathcal{I}_{\mathbf{y}} \in \mathcal{S}\big) \right\}$, where $\mathcal{S} = \Big\{\{4,5,7\}, \{6,4,7\}, \{5,6,7\}, \{1,5,7\}, \{2,6,7\}, \{3,4,7\}\Big\}$. Algorithmically, $\mathcal{S}$ consists of the weight-three error patterns for which message-passing decoding eventually fails to decode, stalling on a ``stopping set''.

\subsection{Mismatch costs of message-passing decoding over the BEC}
\label{subsec:4.B}
We now analyze the mismatch cost of message-passing using the framework of Sec.~\ref{sec:framework}. We quantify how the state distribution of the decoder evolves across each computational step, and how this evolution contributes to the total mismatch cost, parameterized by the channel erasure probability $\epsilon$ and the message prior $\alpha$.
\begin{figure}[!ht]
    \centering 
    \subfloat[Mismatch costs as a function of $\epsilon$.]{
    \includegraphics[width=0.49\textwidth]{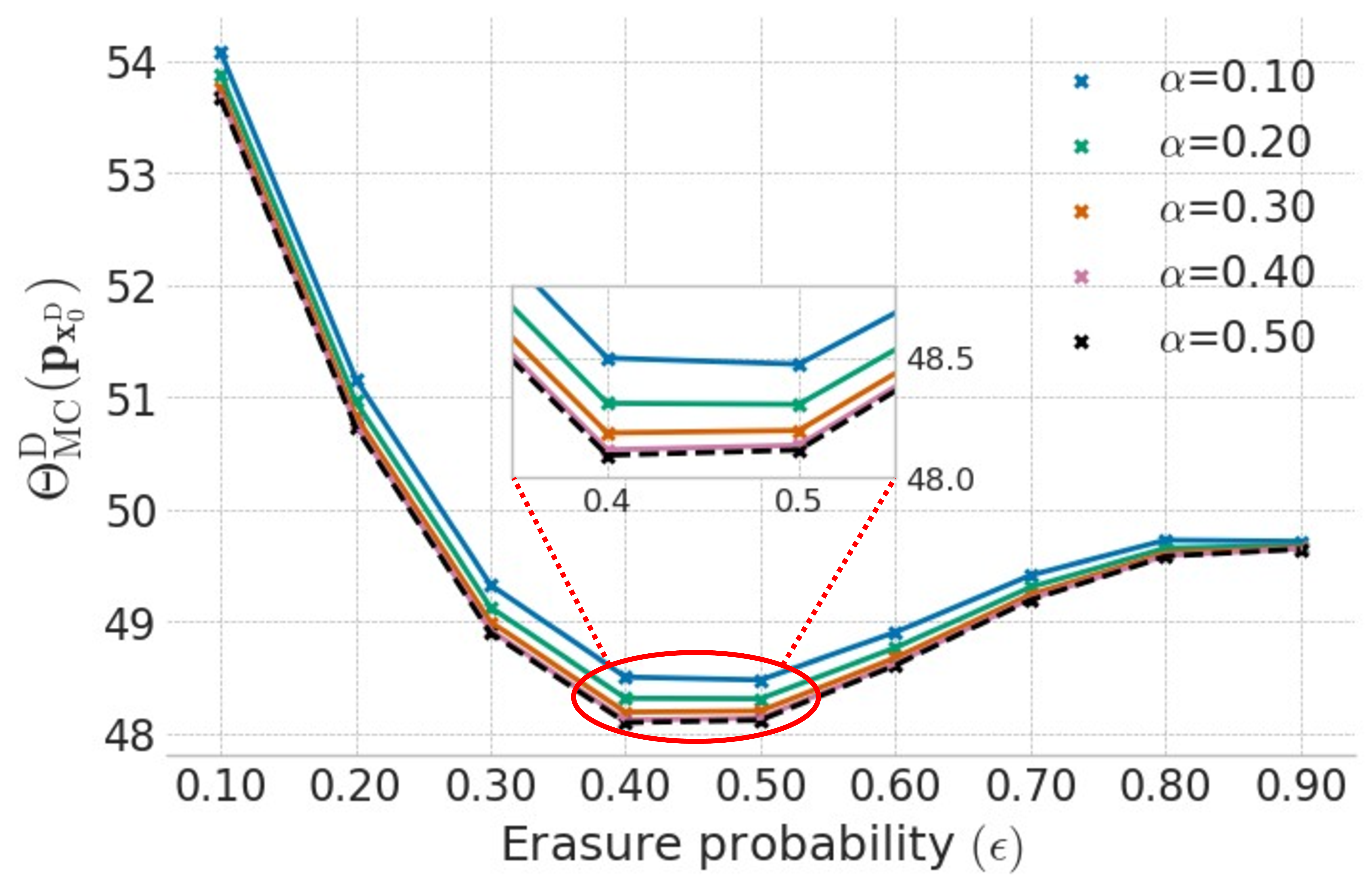}
    \label{subfig:5.1}
    } \hfill
    \subfloat[Mismatch costs as a function of $\alpha$]{
    \includegraphics[width=0.47\textwidth]{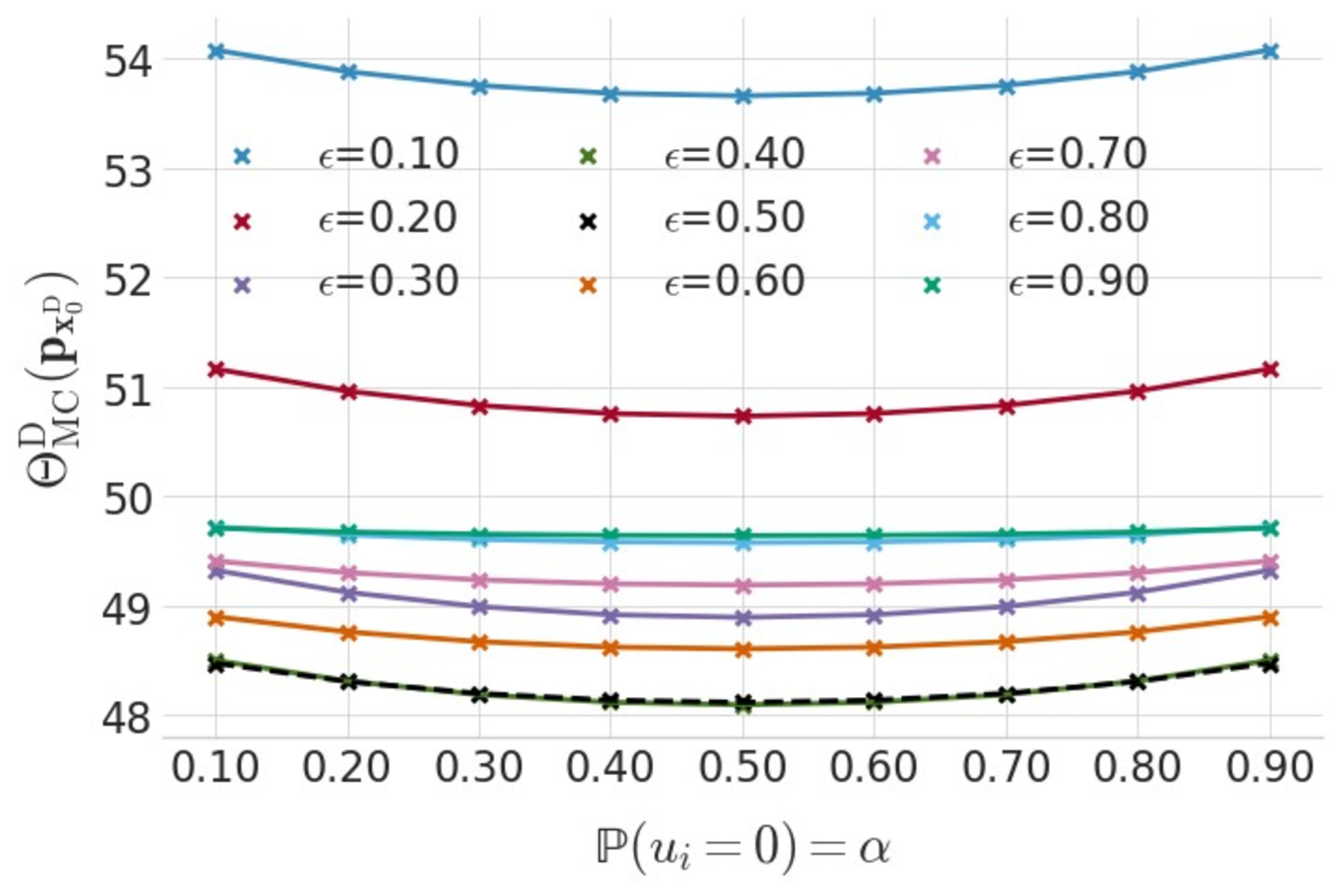}
    \label{subfig:5.2}
    }
    \caption{Mismatch cost of main decoder program $\mmcA{D}{\cdot}$ as $\alpha$ and $\epsilon$ varied.}
    \label{fig:5}
\end{figure}

Figure~\ref{fig:5} plots the mismatch cost $\mmcA{D}{\pbxa{0}{D}}$ as a function of the channel erasure probability $\epsilon$ and message prior $\alpha$. Figure~\ref{subfig:5.1} shows that the mismatch cost exhibits a clear dependence on $\epsilon$.  The dependence is roughly quadratic, with a minimum at $\epsilon = 0.5$. In contrast, given a uniform prior over $\XA{D}$, the mismatch cost shows only minimal variation as $\alpha$ changes.  This is illustrated in Fig.~\ref{subfig:5.2}. The curves are symmetric around $\alpha = 0.5$.  The symmetry follows from the statistical equivalence of input distributions for $\alpha$ and $1-\alpha$, under the uniform prior. In the rest of this section, we will primarily consider $\alpha=0.5$.

\begin{figure}[!ht]
    \centering 
    \subfloat[$\epsilon = 0.1$.]{
    \includegraphics[width=0.31\textwidth]{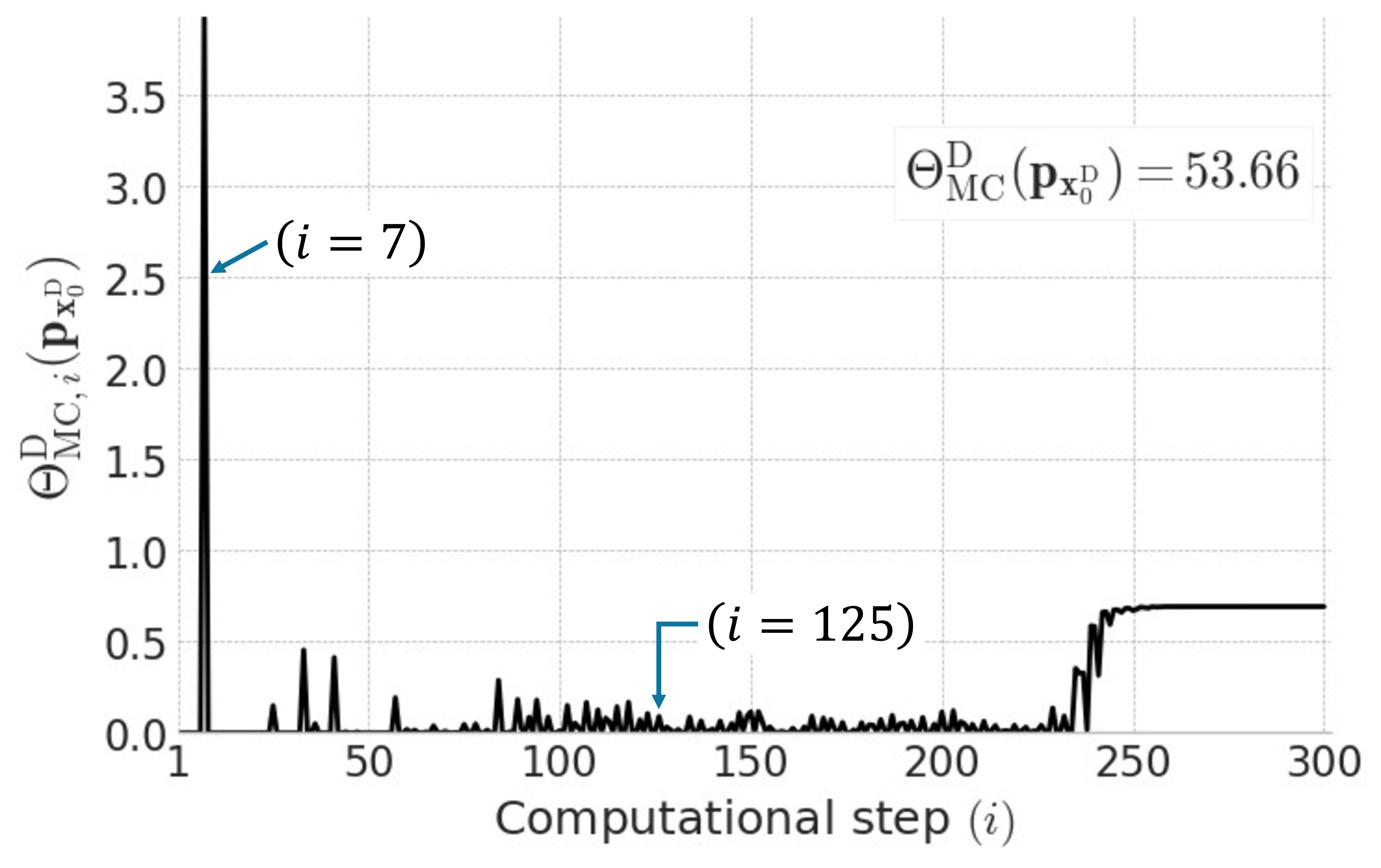}
    \label{subfig:6.1}
    } \hfill
    \subfloat[$\epsilon = 0.5$.]{
    \includegraphics[width=0.31\textwidth]{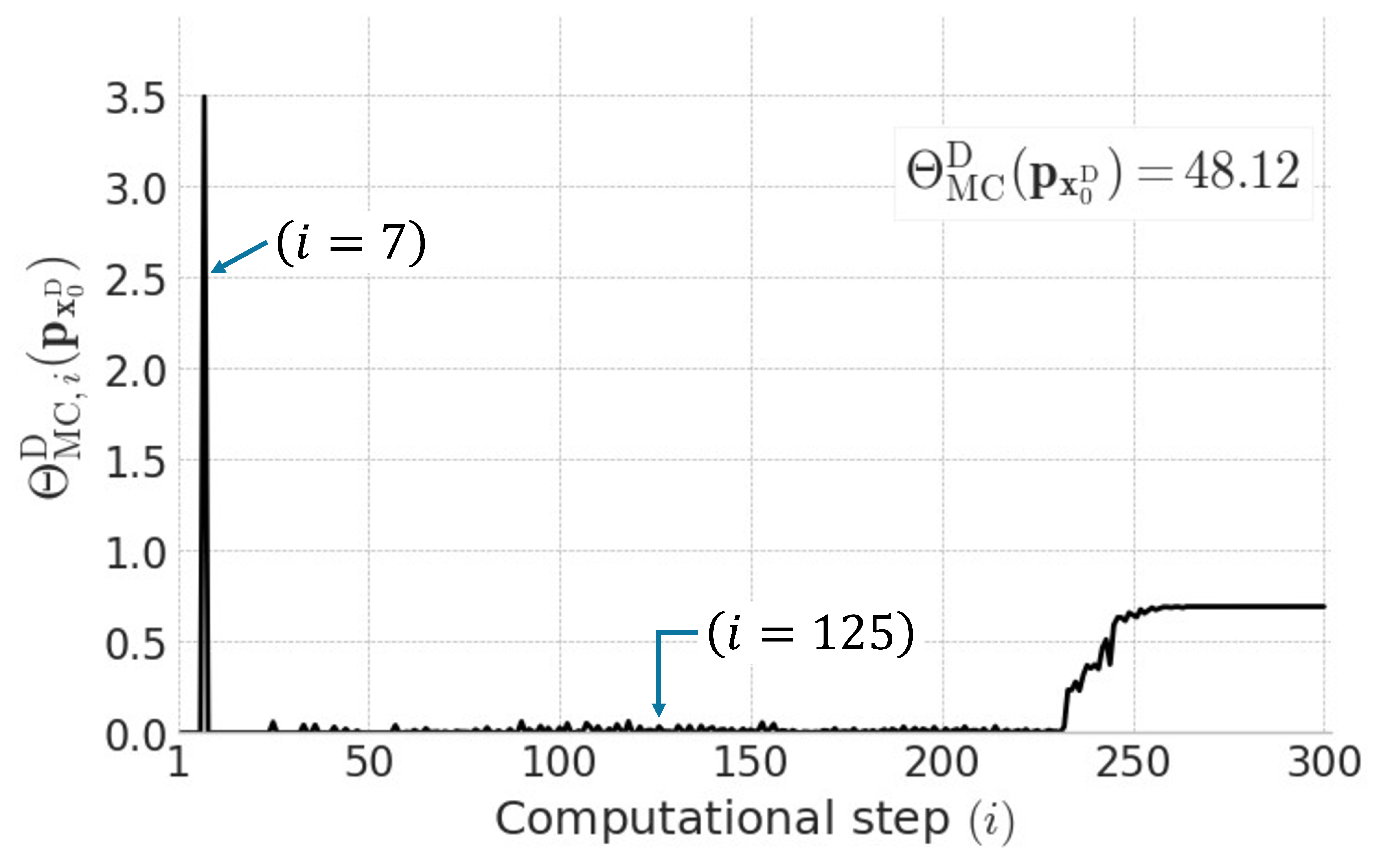}
    \label{subfig:6.2}
    }\hfill
    \subfloat[$\epsilon = 0.9$.]{
    \includegraphics[width=0.31\textwidth]{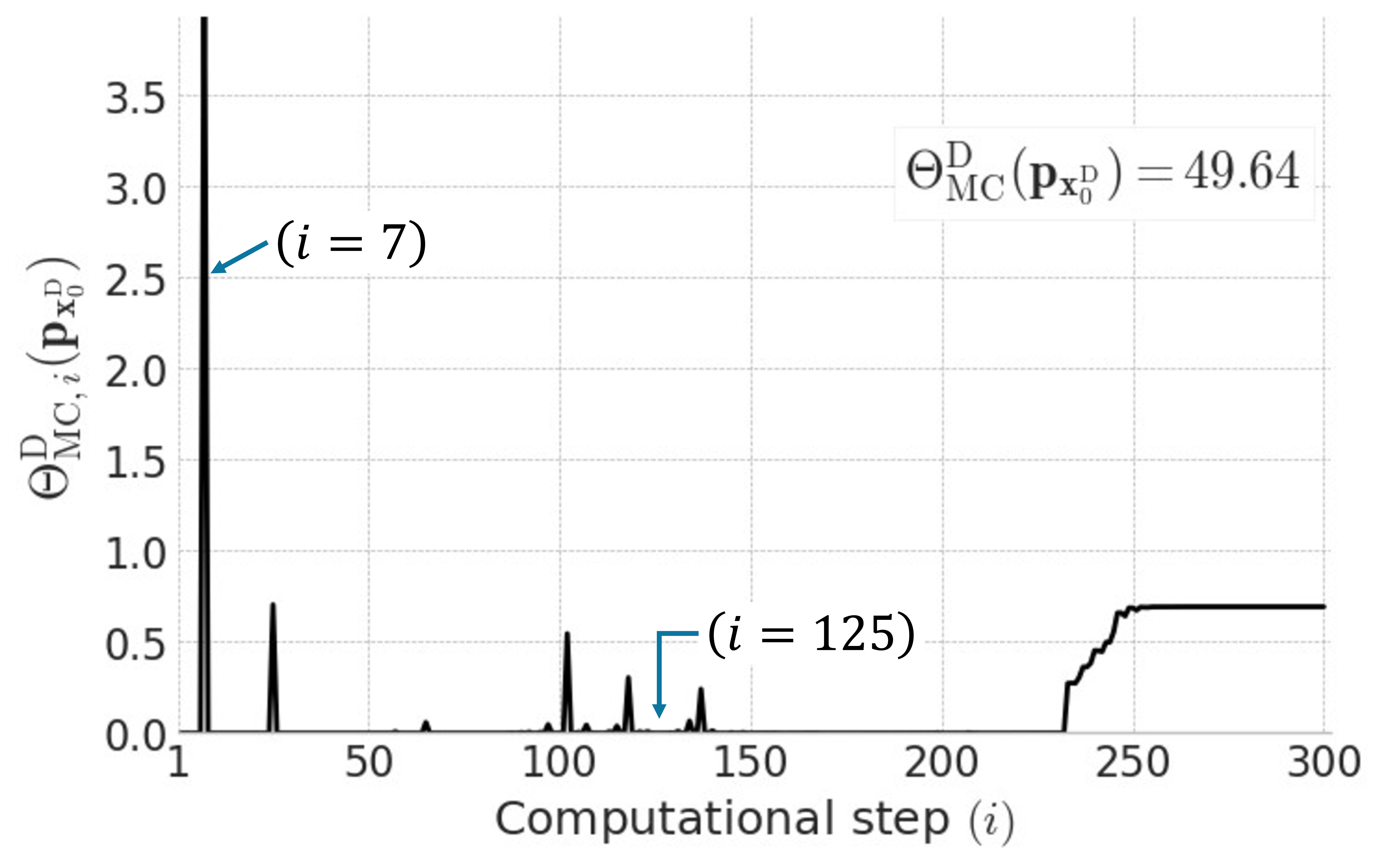}
    \label{subfig:6.3}
    }
    \caption{Step-wise mismatch cost $\pmmcA{D}{\pbxa{0}{D}}$ of the main decoder program for $\alpha = 0.5$ at varying erasure probabilities $\epsilon$. For low $\epsilon$, the mismatch accumulates steadily across many steps, while for high $\epsilon$, it is concentrated in the early steps.}
    \label{fig:6}
\end{figure}

Assuming that the device operates in contact with a thermal bath at $T = 300,\mathrm{K}$, the mismatch cost of $\A{D}$ in Fig.~\ref{fig:5} implies a lower bound on energy dissipation $E(\pxa{0}{D})$ on the order of $0.2\,\mathrm{aJ}$ (attojoules, where $\rm 1\,aJ = 10^{-18}\,J$). However, since Fig.~\ref{fig:5} reports only the total mismatch cost aggregated over all computational steps~\eqref{eq:3.12}, it does not reveal how mismatch cost depends on the decoding process. Therefore, we plot the step-wise mismatch cost for three representative values of $\epsilon$ in Fig.~\ref{fig:6}. This helps to identify which stages of message passing generate the mismatch cost, and how the initial distribution influences such contributions. In all cases, the step-wise mismatch cost stabilizes by $i \approx 250$, indicating the decoder has reached the steady state $\pi_{\gA{D}}$.

The early steps ($i<7$) occur before any message passing. Here, the decoder updates only special registers or advances the program counter $\pc{D}$. The state distribution over $\XA{D}$ is essentially unchanged. The first significant increase in mismatch cost occurs at $i=7$, when messages are read from memory $\A{M}$ into the register $\mr{\mu}$. Before this access ($i=6$), $\mr{\mu}$ may differ across trajectories, so states of the form $\xA{D}_{6} = (\mr{y},\mr{\mu_j},\dots, \pc{D}_l)$ for $j \in [J]$ can all have nonzero probability. The memory read overwrites $\mr{\mu}$ with the unique stored value $\mr{\mu_y}$, mapping these states to $\xA{D}_{7} = (\mr{y},\mr{\mu_y},\ldots, \pc{D}_{l+1})$ and aggregating their probabilities. This many-to-one state compression produces the first spike in mismatch cost. The spike reflects energy irreversibly lost as multiple prior states map to the same current state, making past state evolution unrecoverable. This logical irreversibility contributes to entropy production (cf.~Definition~\ref{def:2}), and by Landauer’s principle implies a lower bound on the associated energy dissipation~\cite{landauer1996minimal}.

For $\epsilon = 0.1$ in Fig.~\ref{subfig:6.1}, the initial probability mass concentrates on states whose $\mr{y}$ corresponds to low-weight, correctable, erasure patterns in $\XA{D}_{\mathcal{E}_{\rm E}}$. During the first message-passing iteration (steps $7 \le i \lessapprox 125$), many of these states map to a smaller set of intermediate states. For example, two initial states differing only in the location of a single erasure, e.g., $(0,1,1,1,1,1,1)$ and $(1,0,1,1,1,1,1)$, both map to the same intermediate state with $\mr{y}^{\ast}=(1,1,1,1,1,1,1)$ after the first iteration. This many-to-one mapping drives the state distribution away from the uniform prior, producing numerous spikes in the step-wise mismatch cost. Subsequent computational steps resolve higher-weight correctable patterns (i.e., weight-$3$ patterns), thereby generating smaller spikes. States corresponding to uncorrectable patterns carry negligible probability and contribute little to the mismatch cost. The cumulative step-wise mismatch cost therefore accounts for the high total mismatch when $\epsilon < 0.3$ in Fig.~\ref{fig:5}, corresponding to an energy dissipation of $0.224\,\rm aJ$.

For $\epsilon = 0.9$ in Fig.~\ref{subfig:6.3}, the initial distribution is dominated by states related to uncorrectable erasure patterns in $\XA{D}_{\mathcal{E}_{\rm H}}$. These states change little, if at all, and contribute minimally to the step-wise mismatch cost. While states related to correctable patterns evolve during $7 \le i \lessapprox 125$, they carry little probability mass. As the states with uncorrectable patterns dominate the overall distribution $\pxa{i}{D}$, the support of the distribution is largely unchanged across steps. Altogether then, the distribution over $\XA{D}$ evolves minimally, producing sparse, high-amplitude spikes and a smaller mismatch cost (i.e., $\sim 0.205\,\rm{aJ}$) than for low $\epsilon$ in Fig.~\ref{subfig:5.1}.

For ${\epsilon = 0.5}$ in Fig.~\ref{subfig:6.2}, all channel outputs $\bm{y} = \smpvin{D}$ occur with probability $2^{-11}$ (per \eqref{eq:3.8}), giving an approximately uniform initial state distribution. The step-wise mismatch cost shows frequent spikes, similar to those at ${\epsilon = 0.1}$ in Fig.~\ref{subfig:6.1}, due to partial collapses of states related to correctable patterns. Compared to both ${\epsilon = 0.1}$ and ${\epsilon = 0.9}$, the spike amplitudes are negligible, as states related to uncorrectable patterns retain sufficient probability mass, keeping the distribution near-uniform. This gives the minimum total mismatch cost for $\A{D}$ (i.e., $0.199\,\rm{aJ}$).

\begin{figure}[!ht]
    \centering 
    \subfloat[Total mismatch cost of message-passing.]{
    \includegraphics[width=0.475\textwidth]{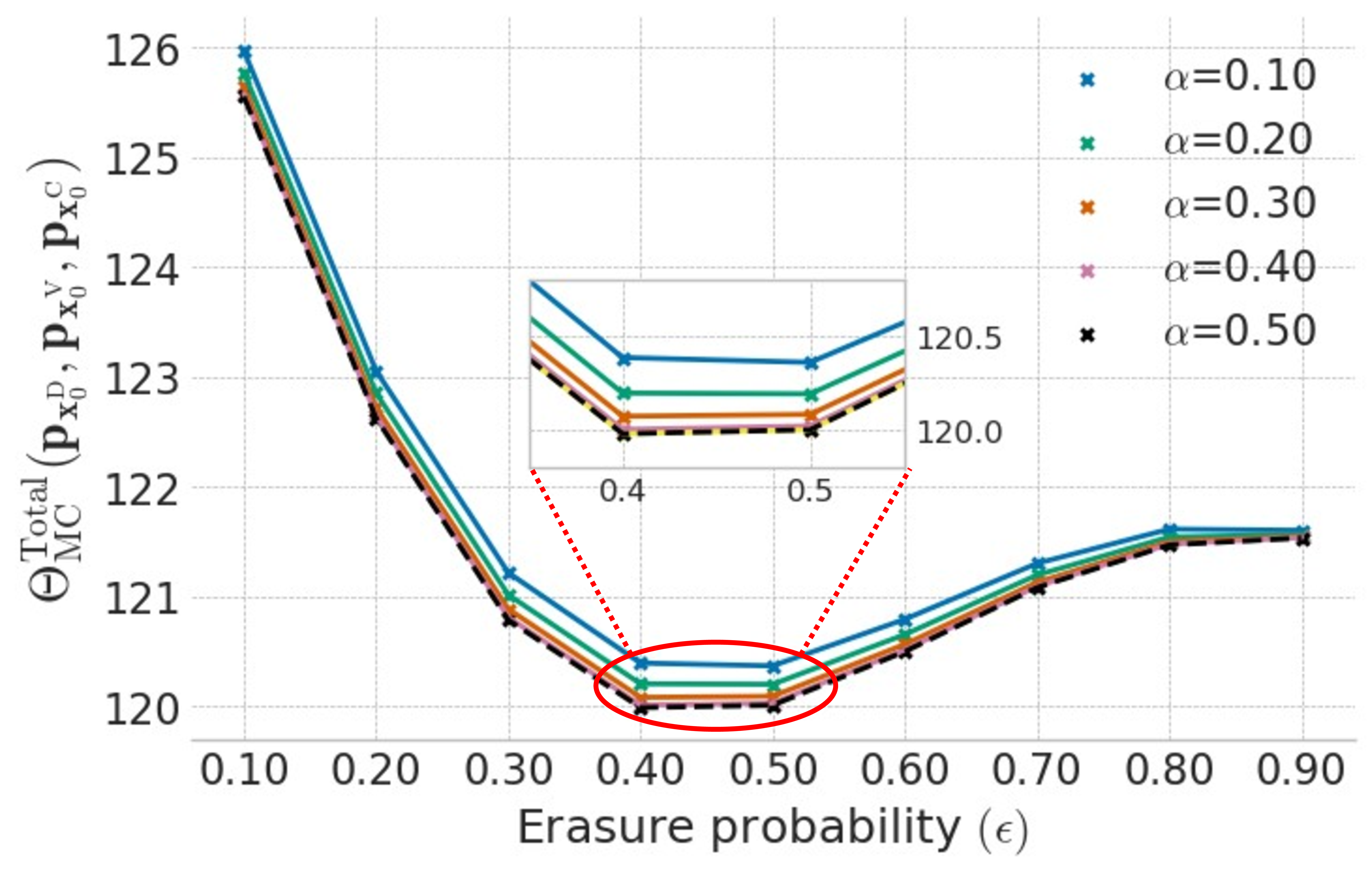}
    \label{subfig:8.1}
    } \hfill
    \subfloat[Decomposition of mismatch costs for $\alpha = 0.5$]{
    \includegraphics[width=0.485\textwidth]{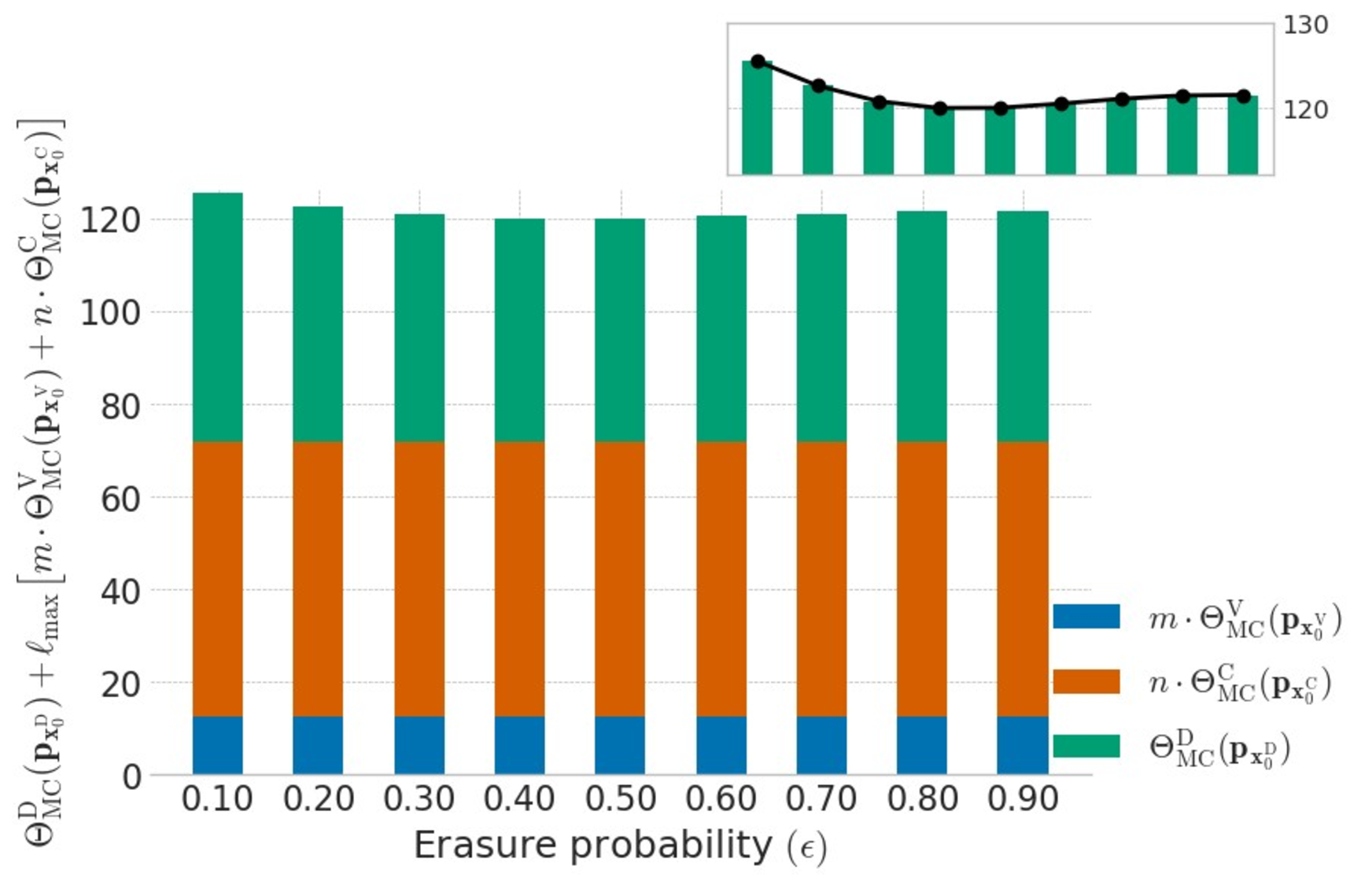}
    \label{subfig:8.2}
    }
    \caption{Total mismatch cost of the message-passing decoder $\mmcA{Total}{\pbxa{0}{D}, \pbxa{0}{V}, \pbxa{0}{C}}$ (per~\eqref{eq:3.19}) and its decomposition into contributions from the decoder $\A{D}$ and subroutines $\A{V}$ and $\A{C}$.}
    \label{fig:8}
\end{figure}

Finally, we consider the total mismatch cost of the message-passing decoder per~\eqref{eq:3.19}, which combines the mismatch costs of the main program $\mmcA{D}{\pbxa{0}{D}}$ and the two subroutines $\mmcA{C}{\pxa{0}{C}}$ and $\mmcA{V}{\pbxa{0}{V}}$ invoked within it. Fig.~\ref{subfig:8.1} shows the total mismatch cost, and the breakdown of the total mismatch cost into its constituent parts (main program, check update, variable updates) for $\alpha = 0.5$ is presented in Fig.~\ref{subfig:8.2}. The corresponding total energy dissipation irreversibly lost to the reservoirs is approximately $0.514\,\rm{aJ}$.

The total mismatch cost shows a roughly convex dependence on $\epsilon$ following $\mmcA{D}{\pbxa{0}{D}}$. The subroutine input distributions are uniform (Sec.~\ref{subsec:3.E}) and independent of $\alpha$ and $\epsilon$. As a result, each subroutine’s contribution to the mismatch cost remains constant (i.e., $0.01\,\rm{aJ}$ per invocation) each time it is called within the main program. The variable-node update dominates the total mismatch cost because it is invoked $n \ell_{\max}$ times (21 in our setting) followed by the main program, with check-node updates contributing the least. The smaller contribution from check nodes arises from their smaller number ($m < n$), even though the per-call mismatch cost $\mmcA{C}{\pbxa{0}{C}}$ is of the same order as that of the variable-node subroutine $\mmcA{V}{\pbxa{0}{V}}$.

The dependence of mismatch cost on $\epsilon$ and $\alpha$ observed above arises from assuming uniform priors for the main program $\A{D}$ and subroutines $\A{V}$ and $\A{C}$. Equivalently, this corresponds to physical implementations with energy functionals $f(\smpxA{A})$ for which the uniform distribution is thermodynamically optimal under~\eqref{eq:2.6}. In practice, however, the hardware and its state-dependent entropy flows are fixed, so ``non-uniform distributions" may instead be thermodynamically optimal and can cause lower mismatch cost. If significant variability in mismatch cost is observed, one could, in principle, select or engineer the prior that minimizes the thermodynamic cost of implementing the decoder. We examine this possibility for $\A{D}$ further next, in Sec.~\ref{subsec:4.D}. 

\subsection{Mismatch cost under a non-uniform prior distribution}
\label{subsec:4.D}
In this section, we analyze how a non-uniform prior over the decoder state space $\XA{D}$ affects the mismatch cost of the main decoder program $\A{D}$. Such priors arise from hardware-dependent thermodynamic characteristics encoded by the energy functional over decoder states. While the computation map $\gA{D}$ remains fixed, the induced prior changes how the mismatch cost depends on $\epsilon$ and $\alpha$, leading to different thermodynamically favorable operating regimes.

\begin{figure}[!ht]
    \centering
    \subfloat[Energy functional $f(\smpxA{D})$ used in optimization.]{
    \includegraphics[width=0.482\textwidth]{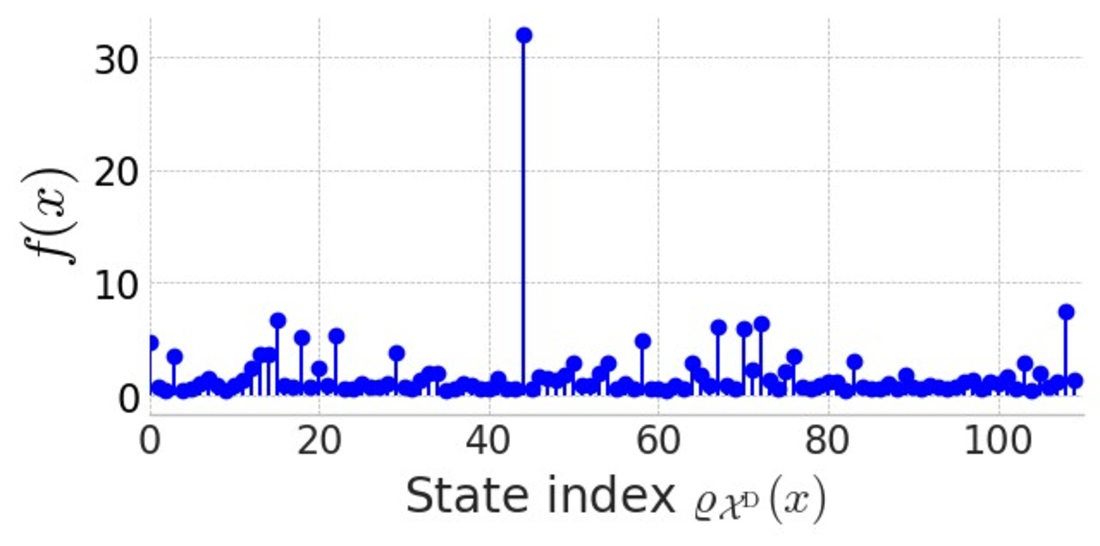}
    \label{subfig:12.1}
    } \hfill
    \subfloat[Corresponding non-uniform prior $\qbtxa{0}{D}$. ]{
    \includegraphics[width=0.482\textwidth]{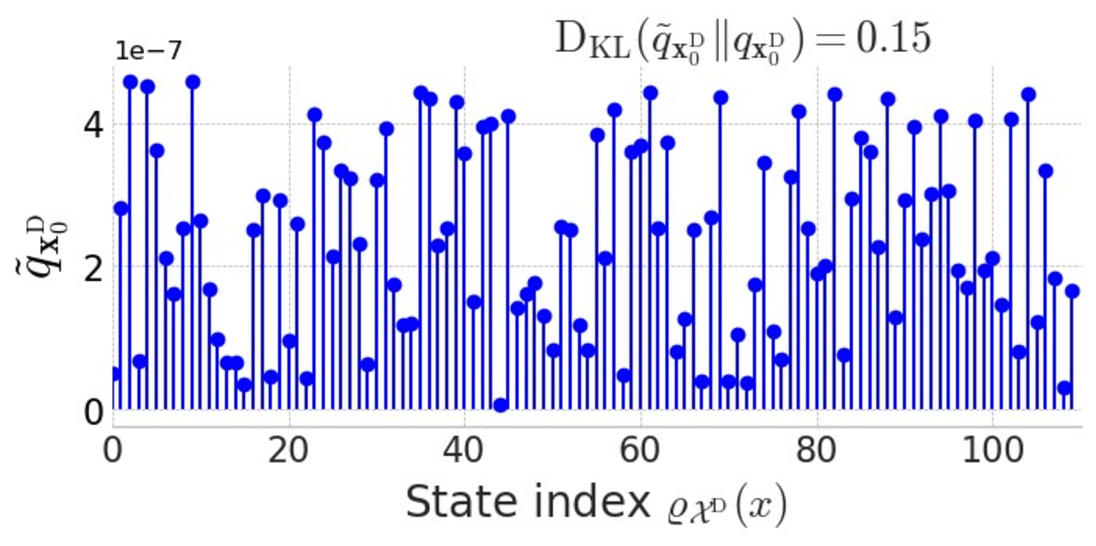}
    \label{subfig:12.2}
    }
    \caption{Constructed energy functional $f(\smpxA{D}) = \beta\Bar{Q}(\smpxA{D})$ and the corresponding prior that minimizes $\EP{\cdot}$. Only a portion of $\XA{D}$ is shown for clarity.
} 
    \label{fig:12}
\end{figure}

To illustrate the effect of non-uniform priors, we construct an entropy-flow functional $f(\smpxA{D}) = \beta \Bar{Q}(\smpxA{D})$ as shown in Fig.~\ref{subfig:12.1}, assigning each state $\smpxA{D} \in \XA{D}$ a distinct average entropy flow. These state-dependent flows represent microscopic hardware-dependent effects underlying the implementation of logical state transitions in $\A{D}$, including variations in electron transition rates, logic switching activity, and leakage currents. However, as discussed in Sec.~\ref{subsec:2.C}, we abstract away from any specific physical implementation. With $\gA{D}$ fixed, we compute the induced prior $\qbtxa{0}{D}$ from~\eqref{eq:2.6} by minimizing entropy production under the prescribed functional $f(\smpxA{D})$. The resulting prior suppresses states with larger entropy flows and favors those with smaller flows, thereby biasing the decoder toward thermodynamically efficient operation (cf.~Fig.~\ref{subfig:12.2}).

\begin{figure}[!ht]
    \centering 
    \subfloat[Mismatch cost as a function $\epsilon$.]{
    \includegraphics[width=0.48\textwidth]{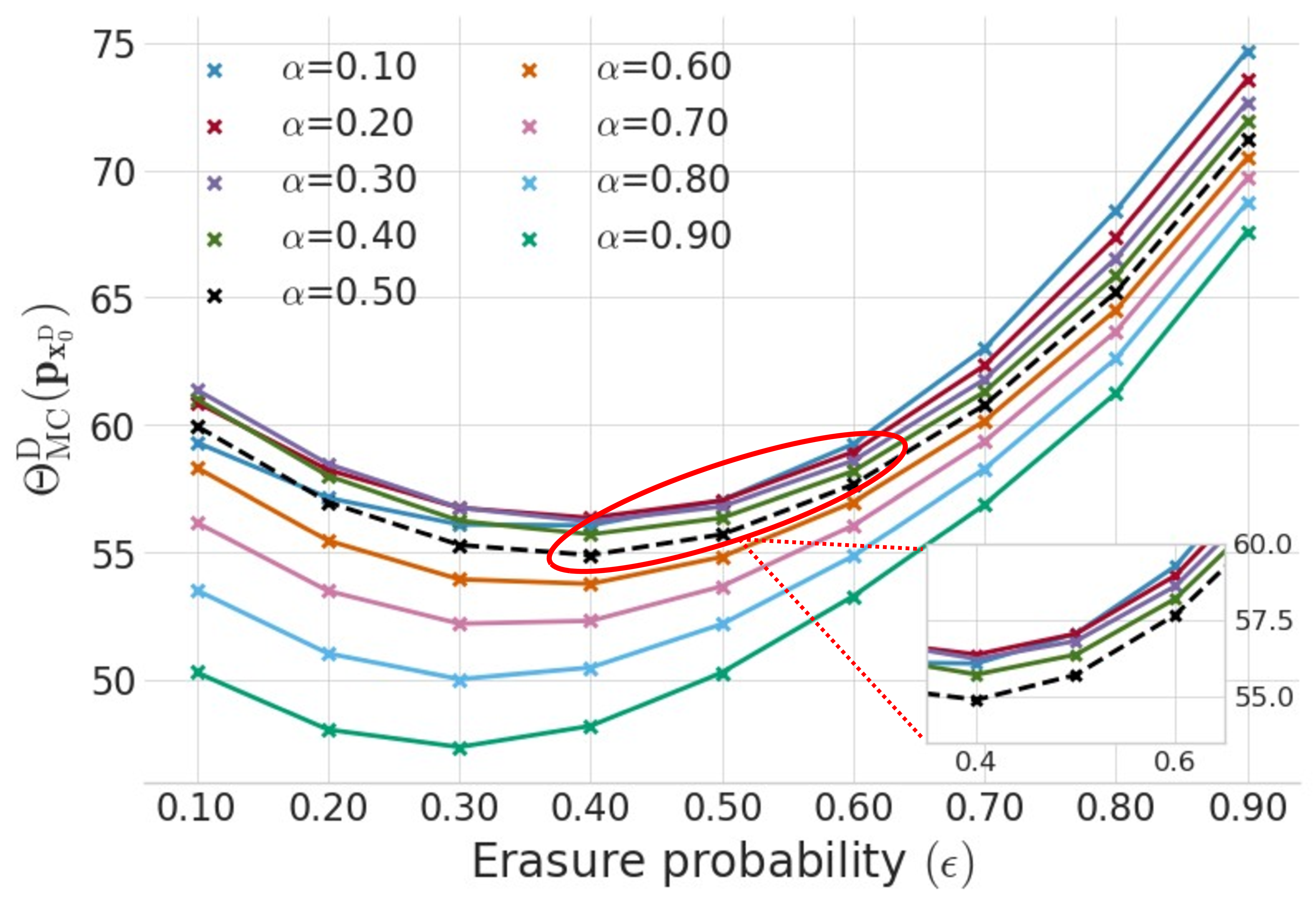}
    \label{subfig:13.1}
    } \hfill
    \subfloat[Mismatch cost as a function of $\alpha$.]{
    \includegraphics[width=0.48\textwidth]{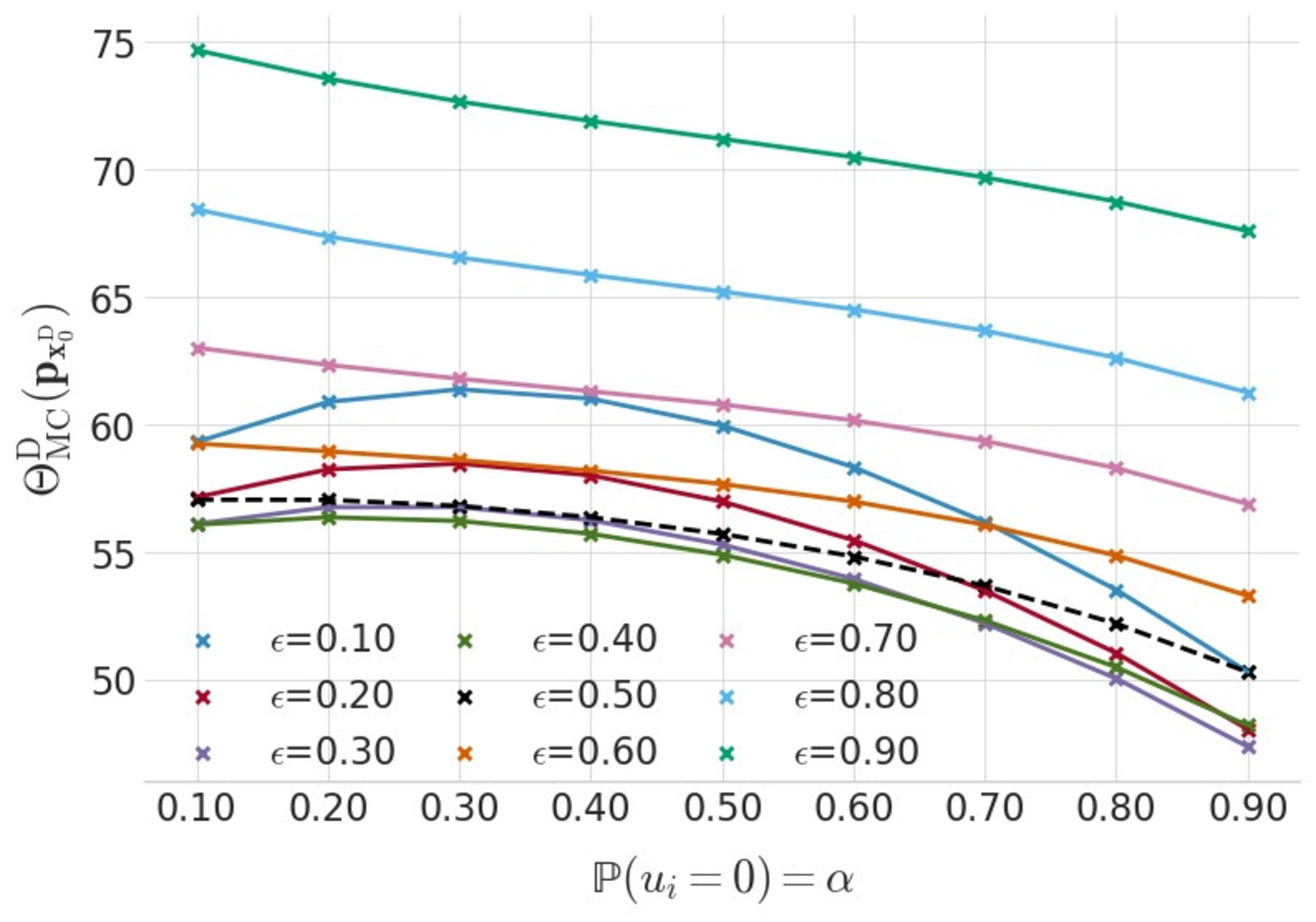}
    \label{subfig:13.2}
    }
    \caption{Mismatch cost of the main decoder program $\A{D}$ under the alternative prior.} 
    \label{fig:13}
\end{figure}

Fig.~\ref{fig:13} shows the mismatch cost of $\A{D}$ as a function of $\epsilon$ and $\alpha$. Under the uniform prior (Sec.~\ref{subsec:4.B}), the mismatch cost is driven primarily by $\epsilon$. Variations in $\alpha$ only weakly affect the cost landscape, with a minimum at $\alpha=0.5$. Under the non-uniform prior $\qbtxa{0}{D}$, this behavior changes: the dependence on $\epsilon$ remains qualitatively unchanged (cf.~Fig.~\ref{subfig:13.1}), while the dependence on $\alpha$ becomes significantly stronger (cf.~Fig.~\ref{subfig:13.2}). Consequently, the minimizing value shifts from $\alpha=0.5$ to $\alpha=0.9$. In addition, for the same initial distribution $\pbxa{0}{D}$, the mismatch cost is consistently larger under the non-uniform prior (cf.~Fig.~\ref{fig:13}) than under the uniform prior (cf.~Fig.~\ref{fig:5}). Specifically, the maximum associated energy dissipation has increased from approximately $0.223\,\mathrm{aJ}$ to $0.31\,\mathrm{aJ}$. This indicates that the prior strongly reshapes how input-distribution parameters influence thermodynamic cost.

This difference in mismatch cost arises from the designed entropy-flow functional $f(\smpxA{D})$ in Fig.~\ref{subfig:12.1}, which assigns distinct entropy flows to decoder states, reflecting microscopic properties of the physical substrate, such as variations in switching activity and leakage currents during computation. This functional induces the prior $\qbtxa{0}{D}$ by favoring low-entropy-flow states and assigning lower probabilities to high-entropy-flow states (cf.~Fig.~\ref{subfig:12.2}). Under the uniform prior, all states in $\XA{D}$ have equal thermodynamic preference at initialization, so trajectory statistics are determined mainly by the intrinsic decoder dynamics encoded by $\gA{D}$. Therefore, the message distribution parametrized by $\alpha$ has only a weak effect on the total mismatch cost. Under the non-uniform prior, initialization is weighted toward low-entropy-flow states. Consequently, $(\alpha,\epsilon)$ determines which decoder states are preferentially visited along trajectories, making the mismatch cost more sensitive to $\alpha$ and shifting the optimal operating point.

Overall, this section considers a physical implementation with a distinct entropy-flow functional $f(\smpxA{D})$, which induces a non-uniform prior over decoder states. Consequently, it was established that even with fixed decoding dynamics defined by $\gA{D}$ and initialization $\pbxa{0}{D}$, thermodynamic efficiency depends on the characteristics of the hardware. This suggests that thermodynamic efficiency can be improved by aligning the prior with the initialization statistics of the computation. In practice, this requires hardware-level properties to be engineered so that low-entropy-flow states emphasized by the prior are preferentially realized during computation.

\subsection{Mismatch costs of correctable and non-correctable erasure patterns}
\label{subsec:4.C}
\begin{figure}[!ht]
    \centering
    \subfloat[Mismatch cost for inputs restricted to ${\bf y} \in  \XAE{D}{E}$. \label{subfig:10.2}]{
    \includegraphics[width=0.48\textwidth]{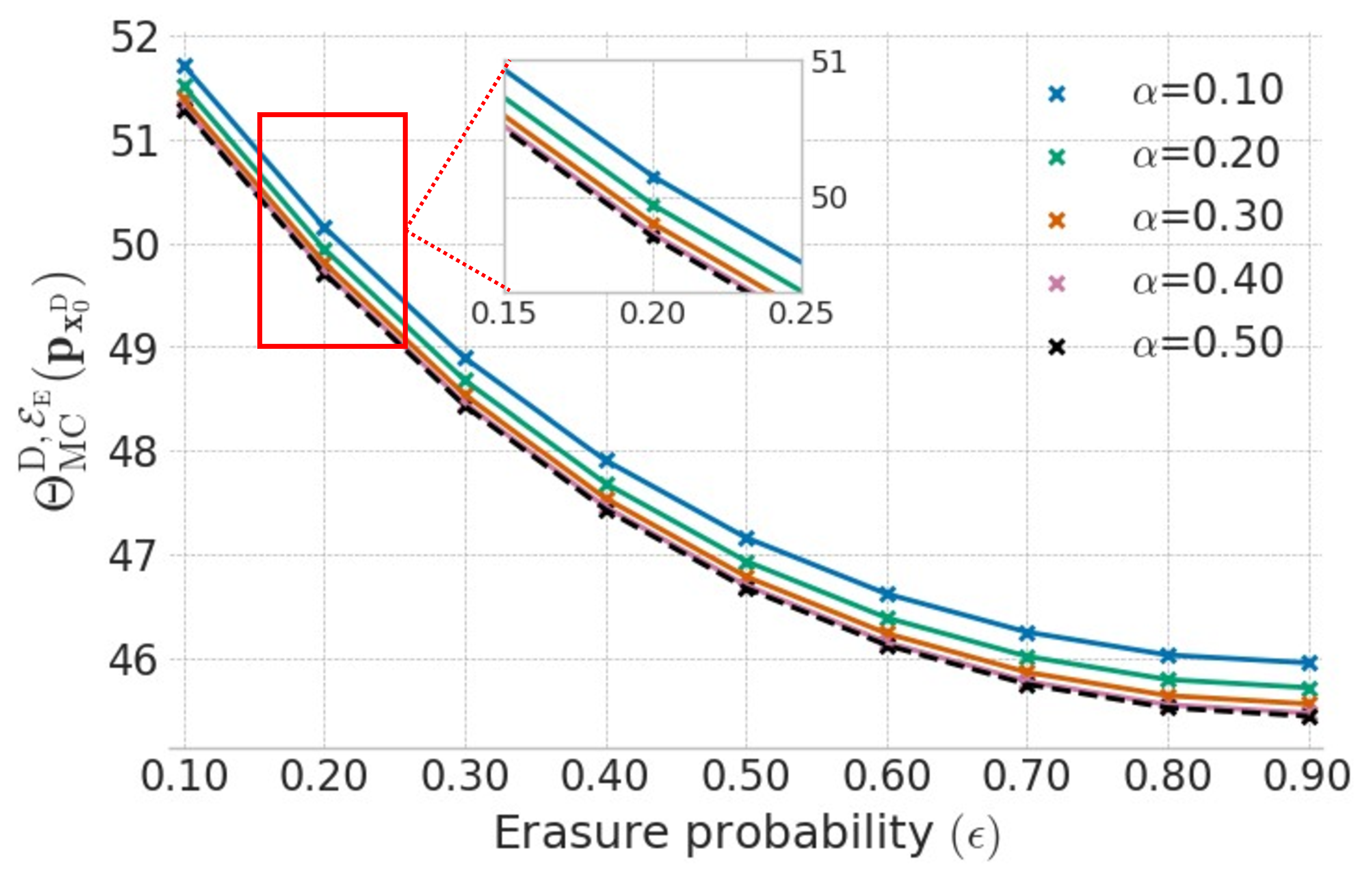}
    }\hfill
    \subfloat[Mismatch cost for inputs restricted to ${\bf y} \in  \XAE{D}{H}$. \label{subfig:10.3}]{
    \includegraphics[width=0.48\textwidth]{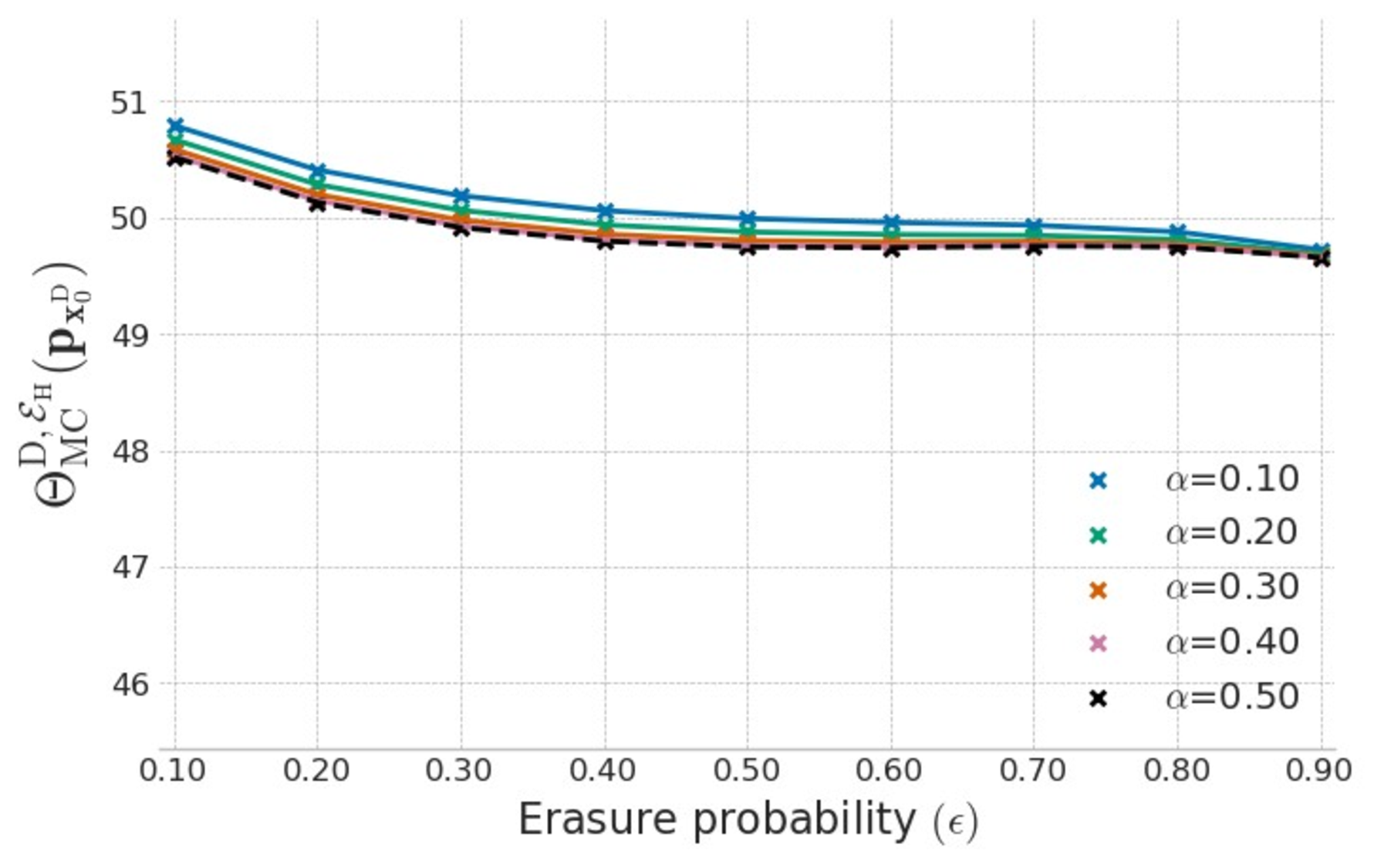}
    
    }
    \caption{Mismatch cost of the main decoder program for input distributions conditioned on specific subsets of erasure patterns.}
    \label{fig:10}
\end{figure}

In this subsection, we return to the uniform prior setting of Sec.~\ref{subsec:4.B} for the main decoder program $\A{D}$ and examine how the mismatch cost changes when the decoder is restricted to processing only correctable or uncorrectable subsets of BEC erasure patterns. Here, the underlying thermodynamic characteristics of the hardware remain unchanged through the same uniform prior, while the actual initialization statistics induced by the restricted input ensemble change. As a result, the mismatch cost becomes highly sensitive to the structure of the input patterns, even under the same prior. The resulting mismatch costs are shown in Fig.~\ref{fig:10}.

\begin{figure}[!ht]
    \centering 
    \subfloat[Input distribution for correctable erasure patterns for $\epsilon = 0.1$]{
    \includegraphics[width=0.48\textwidth]{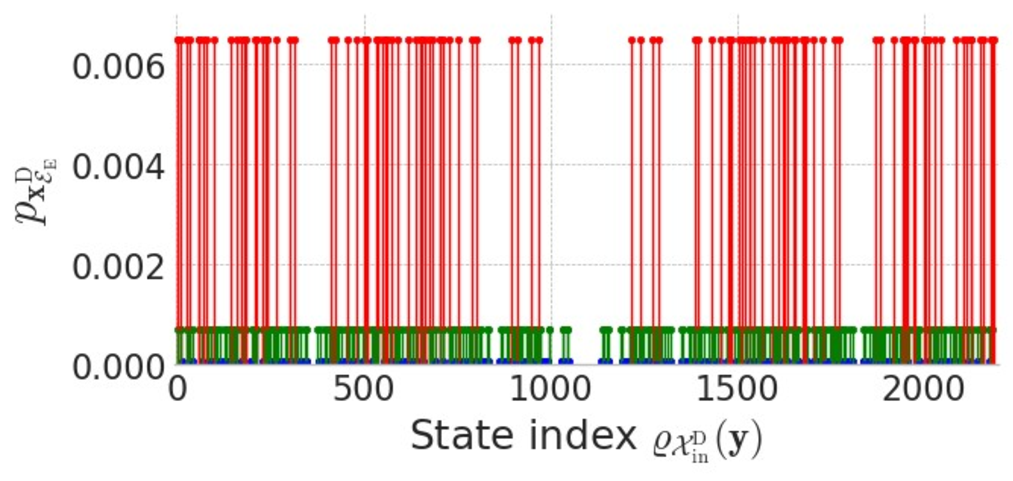}
    \label{subfig:11.3}
    } \hfill
    \subfloat[Step-wise mismatch cost for $\epsilon = 0.1$.]{
    \includegraphics[width=0.48\textwidth]{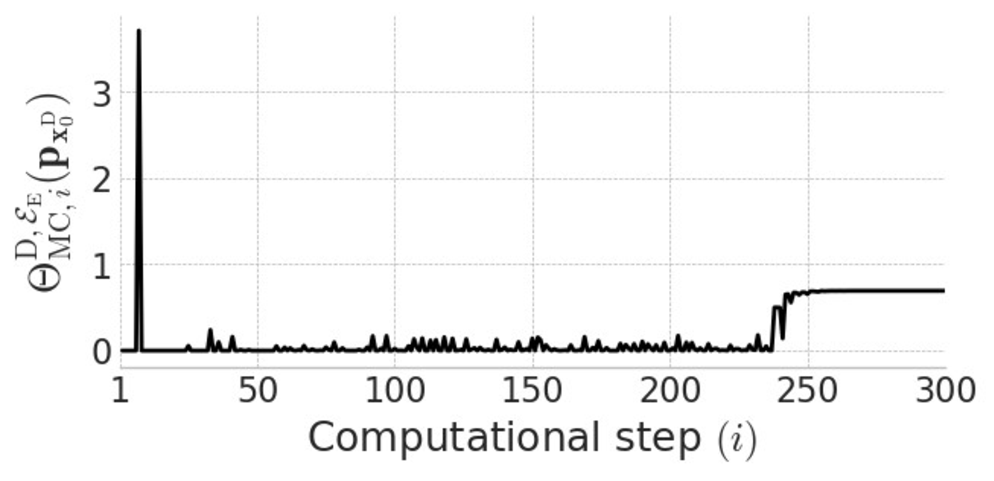}
    \label{subfig:11.1}
    }
    \\
    \subfloat[Input distribution for correctable erasure patterns for $\epsilon = 0.9$]{
    \includegraphics[width=0.48\textwidth]{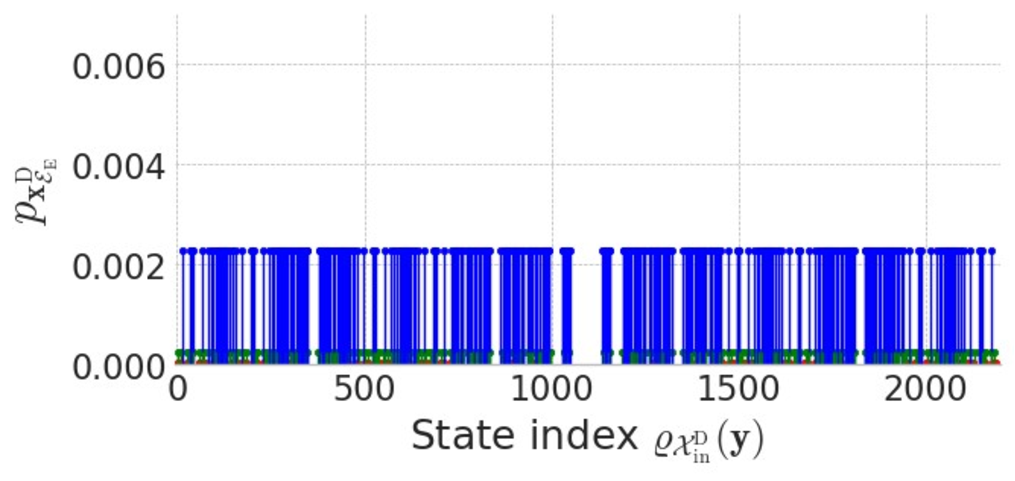}
    \label{subfig:11.4}
    }\hfill
    \subfloat[Step-wise mismatch cost for $\epsilon = 0.9$.]{
    \includegraphics[width=0.48\textwidth]{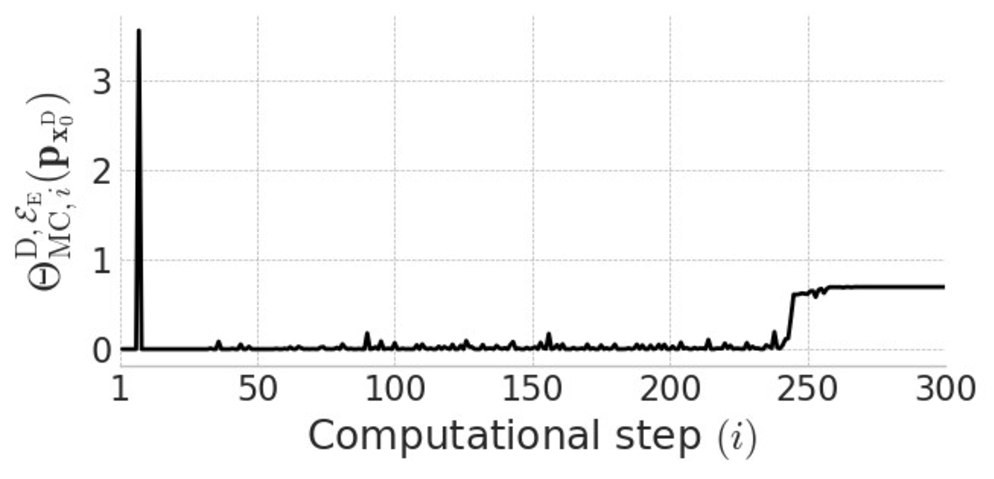}
    \label{subfig:11.2}
    }
    \caption{Input distributions over correctable erasure patterns and corresponding step-wise mismatch costs for $\epsilon=0.1$ and $\epsilon=0.9$ with $\alpha=0.5$. Single-, double-, and triple-erasure probabilities are shown in red, green, and blue, respectively.}
    \label{fig:11}
\end{figure}

For correctable erasure patterns ${\bf y} \in \XAE{D}{E}$, the mismatch cost exhibits a nontrivial dependence on $\epsilon$, as shown in Fig.~\ref{subfig:10.2}. At low erasure probabilities (${\epsilon < 0.5}$), the input distribution over erasure patterns ${\bf p}_{\XAE{D}{E}}$ is dominated by weight-$1$ and weight-$2$ patterns, while weight-$3$ patterns occur with much smaller probability (Fig.~\ref{subfig:11.3}). For example, with $\alpha=0.5$ and $\epsilon=0.1$, a weight-$3$ erasure pattern has a probability $\alpha^4\epsilon^3(1-\epsilon)^4=4\times10^{-5}$, compared to $\alpha^4\epsilon(1-\epsilon)^6=3\times10^{-3}$ for a weight-$1$ pattern. Since single- and double-erasure patterns are corrected after the first message-passing iteration, the induced initial distribution $\pbxa{0}{D}$ in~\eqref{eq:3.21} places significant probability mass on states that map to valid codewords after one iteration. Consequently, many trajectories collapse early, making the decoder state distribution strongly skewed relative to the uniform prior. This produces step-wise mismatch-cost spikes in subsequent computational steps (Fig.~\ref{subfig:11.1}). 
At high erasure probabilities ($\epsilon > 0.5$), weight-$3$ patterns dominate ${\bf p}_{\XAE{D}{E}}$ (Fig.~\ref{subfig:11.4}), while weight-$1$ and weight-$2$ patterns occur with negligible probabilities. Accordingly, $\pbxa{0}{D}$ in~\eqref{eq:3.21} concentrates on a smaller subset of decoder states, so fewer states contribute significantly to the mismatch cost. Since the weight-$3$ erasures need additional decoding iterations, the decoder state distribution evolves more gradually than in the low-$\epsilon$ regime. This reduces both the magnitude and frequency of the mismatch-cost spikes (Fig.~\ref{subfig:11.2}). The result is that the mismatch cost decreases, corresponding to an energy dissipation reduction from $0.215\,\rm aJ$ to $0.186\,\rm aJ$ in Fig.~\ref{subfig:10.2}.

For non-correctable erasure patterns ${\bf y} \in \XAE{D}{H}$, the mismatch cost exhibits a qualitatively different behavior (cf., Fig.~\ref{subfig:10.3}), and remains almost constant for all $\epsilon \in [0,1]$. The non-correctable patterns include weight-$3$ patterns that are only partially resolved, and higher-weight patterns with no recovery. Since none of these patterns can be fully decoded, the decoder state distribution changes only weakly across iterations. At low $\epsilon$ ($<0.5$), partial corrections in the weight-$3$ patterns introduce small localized updates to the state distribution. As $\epsilon$ increases, probability mass shifts toward higher-weight patterns that remain unchanged during decoding, making the decoder dynamics increasingly static. Thus, the induced state evolution does not sufficiently steer the computation toward the low-entropy-flow states favored by the prior. The trajectory ensemble remains largely insensitive to the thermodynamic structure encoded by $f(\smpxA{D})$. This reduces the step-wise variations in mismatch cost. In the limit $\epsilon \to 1$, the all-erased pattern dominates the input distribution, further reducing iteration-to-iteration variation and lowering the overall mismatch cost.

Comparing Figs.~\ref{fig:10} and~\ref{subfig:5.1} shows that conditioning on subsets of erasure patterns reveals thermodynamic structure hidden in the full input distribution. Even under the same uniform prior, the decoder $\A{D}$ exhibits substantially different mismatch costs across input ensembles. This indicates that thermodynamic efficiency depends not only on the decoder dynamics but also on the statistical structure of the inputs that drive the computation. From a hardware perspective, these results suggest that exploiting input statistics can guide the design of more thermodynamically efficient architectures for specialized decoding tasks.

%%%%%%%%%%%%%%%%%%%%%%%%%%%%%%%%%%%%%%%%%%%%%%%%%%%%%%%%%%%%%%%%%%%%%%%%%%%%%%%%%%%%%%%%%%%%%%%%
\section{Conclusion and future work}
\label{sec:futurework}
In this work, we apply the mismatch cost, a thermodynamic metric for periodic machines, to a message-passing decoder modeled as a main program with two subroutines. Memory is treated as an external subsystem. Using the $[7,4]$ Hamming code transmitted over the binary erasure channel as our prototypical example, we show that algorithmic dynamics, input statistics, and prior selection all play a role in developing a lower bound on the energy irreversibly lost in message-passing decoding. 

The $[7,4]$ Hamming code and the BEC enable explicit computation and visualization but have inherent limitations. The short block length and modest error-correcting capability restrict trajectory diversity, while the BEC ignores bit flips and additive noise, limiting relevance to practical channels. Scaling to practical LDPC codes is challenging since the decoder state space grows exponentially with block length, making direct mismatch-cost computation infeasible. Even block-wise sub-matrix methods require storing all trajectories to construct the full transition structure. To address this, future work will explore representative sampling of erasure and noise patterns across different error regimes to estimate mismatch costs for longer codes efficiently. In practice, LDPC performance is typically evaluated using a single transmitted codeword with many induced error patterns; thermodynamic costs can be estimated similarly via sample averages, yielding a tractable lower bound. Finally, we aim to extend the framework to more realistic channels, including the binary symmetric and additive Gaussian channels, using finite-precision representations to maintain tractability. Addressing these limitations will help guide the design of energy-efficient decoding algorithms for practical communication systems.
\ifCLASSOPTIONcaptionsoff
  \newpage
\fi


\end{document}